\documentclass[11pt]{article}
\usepackage{jheppub}

\usepackage{amssymb,amsmath}
\usepackage{amsfonts,amsthm}
\usepackage{tikz}
\usepackage{float}
\usepackage{diagbox}
\usepackage{arydshln}
\usepackage{placeins}

\usepackage{etoolbox}
\appto\appendix{\addtocontents{toc}{\protect\setcounter{tocdepth}{1}}}

\def\be{\begin{eqnarray}}
\def\ee{\end{eqnarray}}
\newcommand{\nn}{\nonumber}

\newcommand{\eqn}[1]{(\ref{#1})}

\newcommand{\bra}[1]{\langle{#1}|}
\newcommand{\ket}[1]{|{#1} \rangle}

\def\Dslash{\,{\raise.15ex\hbox{/}\mkern-12mu D}}
\def\Dbarslash{\,{\raise.15ex\hbox{/}\mkern-12mu {\bar D}}}
\def\delslash{\,{\raise.15ex\hbox{/}\mkern-9mu \partial}}
\def\delbarslash{\,{\raise.15ex\hbox{/}\mkern-9mu {\bar\partial}}}
\def\pslash{\,{\raise.15ex\hbox{/}\mkern-9mu p}}
\def\calDslash{\,{\raise.15ex\hbox{/}\mkern-12mu {\cal D}}}

\newcommand{\Z}{{\bf Z}}

\newcommand{\Tr}{{\rm Tr}}

\def\mes{\displaystyle\int\text{d}^4x\,}

\def\Tr{\text{Tr}\,}
\def\ii{\text{i}}
\def\dd{\text{d}}

\def\Z{\mathbb{Z}}

\def\diag{\text{diag}}

\def\lae{\mathrel{\mathop{\smash{\lower .5 ex \hbox{$\stackrel<\sim$}}}}}
\def\lae{\mathrel{\mathop{\smash{\lower .5 ex \hbox{$\stackrel>\sim$}}}}}

\newcommand{\Ztwo}{\ensuremath{{\bf Z}_2}}
\def\Z2{\Ztwo}

\def\lstrong{\Lambda_{\rm strong}}
\def\lweak{\Lambda_{\rm weak}}

\title{If the Weak Were Strong and  the Strong Were Weak}

\author[1]{Nakarin Lohitsiri}
\author[1,2]{, David Tong}

\affiliation[1]{Department of Applied Mathematics and Theoretical Physics, \\ University of Cambridge, Cambridge, CB3 OWA, UK}
\affiliation[2]{School of Physics, Korea Institute for Advanced Study \\ 
85 Hoegi-ro Dongdaemun-gu, Seoul 02455, Korea}

\abstract{We explore the phase structure of the Standard Model as the relative strengths of the $SU(2)$  weak force and $SU(3)$ strong force are varied. With a single generation of fermions, the structure of chiral symmetry breaking suggests that there is no  phase transition as we interpolate between the $SU(3)$-confining phase and the $SU(2)$-confining phase.  Remarkably, the massless left-handed neutrino, familiar in our world, morphs smoothly into a massless right-handed down-quark. With multiple generations, a  similar metamorphosis occurs, but now  proceeding via a phase transition.

In the second half of the paper we introduce a two-parameter extension of the Standard Model, a chiral gauge theory with gauge group $U(1)\times Sp(r)\times SU(N)$. We again explore the phase structure of the theory as the relative strengths of the $Sp(r)$ and $SU(N)$ gauge couplings vary. }
\begin{document}

\maketitle
\vspace{-2em}

\section{Introduction}

There are two  energy scales in the Standard Model. In combination with a handful of dimensionless couplings and some simple, yet intricate, dynamics, these give birth to the  wondrous diversity of  scales, spanning many orders of magnitude, that emerge  in  nuclear physics, atomic physics and condensed matter physics. 

The two scales are the Higgs expectation value $v$, and the scale $\Lambda_{\rm strong}$, usually called $\Lambda_{QCD}$, at which the strong force lives up to its name. They take values
\be v\approx 250 \ {\rm GeV}\ \ \ {\rm and}\ \ \ \Lambda_{\rm strong} \approx 250 \ {\rm MeV}\nn\ee
There is also a third, counterfactual scale in the Standard Model, which doesn't play any role in our world. This is the infra-red scale at which the weak force would become strong if other effects didn't first intervene. It is a rather academic exercise to specify this scale but if we were to ignore electroweak symmetry breaking and run the $SU(2)$ beta function down, assuming a single massless generation, it is given by\footnote{Obviously, if the Higgs mechanism turns off then all three generations become massless, together with any further generations that lie beyond our current reach. This slows the running of the beta functions. With three massless generations, and the dimensionless couplings fixed to their values at $80\ {\rm GeV}$, we have $\Lambda_{\rm strong} \approx  40 \ {\rm MeV}$ and  $\Lambda_{\rm weak} \approx 2\times 10^{-15}\ {\rm eV}$.}
\be \Lambda_{\rm weak} \approx 3 \times 10^{-3} \ {\rm eV}\nn\ee
The purpose of this paper is to understand the phase structure and possible quantum phase transitions of the theory as the three scales, $v$, $\Lambda_{\rm strong}$ and $\Lambda_{\rm weak}$,  vary.

The question of what happens if the Higgs mechanism is turned off, and the strong force dominates, has been well studied. This situation occurs in the regime,
\be v\ll \Lambda_{\rm weak} \ll \Lambda_{\rm strong}\label{regime1}\ee
It was pointed out long ago that the chiral condensate of QCD transforms under electroweak symmetry. This means that  the pions act as a substitute for the Higgs boson, giving a mass of order $f_\pi$ to the W- and Z-bosons, an observation that motivated the subsequent development of technicolor models \cite{wein2,susskind}. The phenomenology of this  regime was described in \cite{sam} and, in more detail, by Quigg and Shrock  \cite{quigg}.

In this paper, we are interested in what happens as we vary the couplings, to interpolate from \eqn{regime1} to the regime
\be v\ll \Lambda_{\rm strong} \ll \Lambda_{\rm weak}\label{regime2}\ee
The pattern of  chiral symmetry breaking in this regime was mentioned briefly in \cite{quigg} and explored further in  \cite{shrock} and will be reviewed in some detail below. First the weak force with $SU(2)$ gauge group  confines, with a particular pattern of chiral symmetry breaking. This condensate breaks the strong gauge group, $SU(3) \rightarrow SU(2)$, which itself subsequently confines, breaking chiral symmetry yet further.   The resulting physics  shares some similarities with the early work of Abbott and Farhi  \cite{af,af2,af3}, exploring the possibility that the $SU(2)$ weak  force is actually confining, rather than spontaneously broken.

Our goal is to understand the spectrum of massless fermions and Goldstone bosons of the Standard Model, and a closely related chiral gauge theory, in  the two regimes \eqn{regime1} and \eqn{regime2}. Our primary motivation for undertaking this work is simple: we thought it was a fun question. More generally, this paper sits within a larger programme aimed at understanding the dynamics of chiral gauge theories. Early work on this topic is summarised in the review article \cite{ball}.  Since then, a number of articles have studied the dynamics and phase structure of large classes of chiral gauge theories \cite{appel0,appel2,appel3,ss1,ss2,ss3,stefano,ryttov,stefano2,anber}. A number of proposals have been made for lattice regularisations of chiral gauge theories  \cite{lattice1,lattice2,lattice3,lattice4,lattice5}.

\subsubsection*{Summary of Results}

As we  vary the coupling constants of the theory, to interpolate from regime \eqn{regime1} to regime \eqn{regime2}, the physics depends strongly on the number of generations, which we denote as $N_f$. 

Perhaps the biggest surprise arises when we have just a single generation, $N_f=1$. In this case, the massless fermion spectrum is comprised of a single, left-handed neutrino in the regime \eqn{regime1}, a fact that is familiar from our world\footnote{This statement holds in the absence of a right-handed neutrino. However, as we describe in the bulk of the paper, the essential physics remains unchanged by the addition of  a right-handed neutrino.}. In contrast, in regime \eqn{regime2} the massless fermion spectrum contains a single colour component of the right-handed down quark. 
Furthermore, the unbroken symmetries are identical in the two regimes. In particular, the massless down quark that emerges when the weak force dominates is neutral under electromagnetism, a fact that can be understood by noting that the $U(1)_Q$ electromagnetic subgroup twists within the $SU(2)\times SU(3)$ gauge group as we vary the ratio $\Lambda_{\rm strong}/\Lambda_{\rm weak}$.
 These facts suggest that the two regimes sit in the same phase, and the neutrino morphs smoothly into the down quark\footnote{This conclusion is based only on the breaking pattern of the continuous global symmetries. It may well be that more subtle symmetries, such as the higher form symmetries described in \cite{higher1,higher2} give a finer classification of the phases. These ideas were applied to bi-fundamental,  but non-chiral, gauge theories in \cite{tanizaki,karasik}. We hope to return to this question in the future}.

With $N_f\geq 2$ generations, there is a similar story: in regime \eqn{regime1}, one finds massless left-handed leptons, while in regime \eqn{regime2} there are massless, neutral right-handed quarks. This time, however, the symmetry breaking structure in the two regimes differs, ensuring that there is a phase transition between the two. The exact structure of the symmetry breaking depends on the fields  and couplings that are present, and we consider a number of variations of the Standard Model, both with and without hypercharge and Yukawa couplings.

In Section \ref{itsnewsec}, we introduce a novel chiral gauge theory, based on the gauge group
\be G= U(1)_Y \times Sp(r) \times SU(N)\nn\ee
When coupled to specific fermion and scalar fields, this can be thought of as a two parameter extension of the Standard Model. We again explore the phase structure as the relative couplings of the two non-Abelian gauge groups are varied and find a pattern analogous to that of the Standard Model.

Finally, we include two  extended Appendices which describe a number of features of {\it vacuum alignment}, the dynamical process that determines the vacuum structure of theories with chiral symmetry breaking and multiple gauge groups \cite{peskin,preskill}. This, it turns out,  is important in order to understand the structure of chiral symmetry breaking in regime \eqn{regime2}.

\subsubsection*{Note Added}

As we started to write this paper, we became aware of  the  \cite{scooped} by Berger, Long and Turner which asks essentially the same questions, motivated by early universe baryogenesis. Our results largely agree where there is overlap.

\section{Variations on the Standard Model}\label{nonabsec}

We start by discussing a simple chiral gauge theory with gauge group
\be G = SU(2) \times SU(3)\nn\ee
We stick to convention and refer to $SU(2)$ as the {\it weak force} and $SU(3)$ as the {\it strong force}. However, we will be interested both in situations where these names are appropriate, and also in  situations where the weak is strong and the strong is weak. We will encounter a large number of different groups below, both gauge and global; where there is a possibility of confusion, we will refer to the gauge groups as $SU(2)_{\rm weak}$ and $SU(3)_{\rm strong}$.

We couple four Weyl fermions to $G$: these are the left-handed quarks $q_L$, left-handed leptons $l_L$, and right-handed quarks $q_R=(d_R,u_R)$. 
%
%
At this stage, we  include neither the right-handed electron, nor right-handed neutrino since both are singlets under $G$. These  will be introduced in Sections \ref{hypersec} and \ref{itsnewsec} respectively. 

We start by considering just a single generation of fermions; we then turn to multiple generations in Section \ref{nfmore}. 
The non-anomalous global symmetry is
\be F = SU(2)_R \times U(1)_V\label{fuv}\ee
where here, and elsewhere, we ignore discrete factors. Here $U(1)_V$ is the familiar $B-L$ symmetry of the Standard Model.  The transformations of the various fermions under the gauge and global symmetries are summarised as
\FloatBarrier
\begin{table}[h!]
  \begin{center}
    \begin{tabular}{c||c:c|c:c} 
    & \multicolumn{2}{c|}{$G$}  &  \multicolumn{2}{c}{\ \ $F$}  \\
    
  & $SU(2)$  & $SU(3)$ & $SU(2)_R$ & $U(1)_V$ \\
     \hline
     $q_L$ & ${\bf 2}$ & ${\bf 3}$ & ${\bf 1}$ & $+1$ \\
      $l_L$ & ${\bf 2}$ & ${\bf 1}$ & ${\bf 1}$ & $-3$  \\
       $q_R$ & ${\bf 1}$ & ${\bf 3}$ & ${\bf 2}$ & $+1$ 
    \end{tabular}
  \end{center}
\end{table}
\FloatBarrier
\noindent
Both baryon and lepton number are anomalous, meaning that these quantum numbers are not individually conserved in the quantum theory. We will see a dramatic illustration of this fact as we adiabatically vary the coupling constants.

There are 't Hooft anomalies  for both $U(1)_V^3$ and $SU(2)_R$ \cite{thooft}. (The latter is a ${\bf Z}_2$ anomaly \cite{wittensu2}.) This ensures 
that either these symmetries are spontaneously broken in  the infra-red, or there are massless fermions. We will find that the anomalies are saturated in the infra-red by massless fermions, albeit with different microscopic representatives playing the role in different regimes. 

Both gauge group factors are asymptotically free. This means that the  gauge couplings become large in the infra-red where, left to the their own devices, they result in two dynamically generated scales. We refer to the scales for $SU(2)$ and $SU(3)$ as $\lweak$ and $\lstrong$ respectively. We have little understanding of the dynamics when $\lstrong\approx \lweak$. However, in the two limits $\lstrong\gg\lweak$ and $\lweak \gg \lstrong$, some simple intuition about confinement and chiral symmetry breaking is enough to understand what happens. We will then try to match the two regimes. 

\subsubsection*{\underline{$\lstrong\gg \lweak$}}

The limit where the strong force dominates is well studied  \cite{wein2,susskind,sam,quigg}. The strong dynamics results in a quark condensate which takes the form
\be \langle q_{L\,i}^\dagger q_{R\,j}\rangle \sim \lstrong^3\delta_{ij}  \ \ \ \ i,j=1,2\nn\ee
If we ignore the weak force, then this is a condensate in  two-flavour QCD. The $SU(2)_L\times SU(2)_R$ flavour symmetry is spontaneously broken to the diagonal subgroup $SU(2)_{\rm diag}$, resulting in three Goldstone bosons.

Turning on the weak force, we identify $SU(2)_L$ with the $SU(2)_{\rm weak}\subset G$ gauge group. The condensate acts as  a Higgs field, completely breaking the $SU(2)$ gauge group. All three of the would-be Goldstone bosons are eaten, giving mass to the W-bosons. This mass is of order $f_\pi$, the pion decay constant.

A global symmetry survives, formed from a diagonal combination $SU(2)_{\rm diag} \subset SU(2)_{\rm weak} \times SU(2)_R$, and the infra-red global symmetry takes the same form as the ultra-violet symmetry \eqn{fuv},
\be F_{\rm strong} = SU(2)_{\rm diag} \times U(1)_V\nn\ee
The quark bound states are all massive and form representations of $F_{\rm strong}$. Meanwhile, the leptons $l_L$ remain massless, transforming as
\FloatBarrier
\begin{table}[h!]
  \begin{center}
    \begin{tabular}{c||c:c}     
  &  $SU(2)_{\rm diag}$ & $U(1)_V$ \\
     \hline
        $l_L$  & ${\bf 2}$ & $-3$  \\
    \end{tabular}
  \end{center}
\end{table}
\FloatBarrier
\noindent
These massless leptons saturate the 't Hooft anomalies of the global symmetry.

\subsubsection*{\underline{$\lweak\gg \lstrong$}}

When the weak force dominates, we expect a condensate of left-handed fermions to form. There are four such left-handed fermions, each a doublet of $SU(2)_{\rm weak}$. We write these as 
\be \psi_{m} = (q_{L,1},q_{L,2},q_{L,3},l_L)\ \ \ \ m=1,2,3,4\label{psim}\ee
The condensate takes the general form
\be \langle \epsilon_{\alpha\beta}\psi_m^\alpha \psi_n^\beta\rangle\sim \lweak^3 J_{mn}\label{weakcond}\ee
where $\alpha,\beta=1,2$ are $SU(2)_{\rm weak}$ indices, and $J_{mn}$ is a $4\times 4$ anti-symmetric matrix.

If we ignore the strong force, then the $SU(2)$ gauge theory enjoys an $SU(4)$ global symmetry, under which $\psi$ transforms in the ${\bf 4}$. The condensate breaks\footnote{We use the convention $Sp(1)\equiv SU(2)$.} this to $Sp(2)$, resulting in $\dim SU(4) - \dim Sp(2) = 15-10 = 5$ Goldstone bosons. 

Now we turn the strong force back on, and see the effect of the condensate \eqn{weakcond}. This was discussed previously in \cite{shrock}. It turns out that the choice of $J_{mn}$ does not affect the physics in this case. (This statement no longer holds when we discuss multiple generations; we will discuss this in more detail in Section \ref{nfmore} and in much more detail in the Appendices.)  For any choice of $J_{mn}$,  the condensate \eqn{weakcond} includes a quark-bilinear of the form
\be \langle  q_{L\,a} \cdot q_{L\,b}\rangle \sim \lweak^3\epsilon_{abc}\sigma^c\label{colcond}\ee
where $a,b,c=1,2,3$ are $SU(3)_{\rm strong}$ indices and we've now suppressed the $SU(2)_{\rm weak}$ index structure which remains as in \eqn{weakcond}.  For any  choice of  $\sigma^c$, the condensate acts as Higgs field for the strong force, breaking 
\be SU(3)_{\rm strong} \rightarrow SU(2)_{\rm strong}\nn\ee
All 5 of the would-be Goldstone bosons described above are eaten by the now-massive gluons.

Without loss of generality, we choose $\sigma^c=(0,0,1)$, so that the condensate \eqn{colcond} involves only the $q_a$ with $a=1,2$ coloured quarks. We denote the remaining quark as $\hat{q}_L = q_{L\,3}$. It forms a condensate with the lepton
\be \langle  \hat{q}_L\cdot l_L\rangle \sim \lweak^3\label{colweakcond}\ee
where the anti-symmetry of \eqn{weakcond} is assured because the condensate is symmetrised over both spinor and weak indices, leaving the Grassmann nature of the fermions to do its job.

Both condensates \eqn{colcond} and \eqn{colweakcond} would appear to break the $U(1)_V$ symmetry of the original theory; they have charges $+2$ and $-2$ respectively.  However, it is straightforward to find a global $U(1)$ symmetry that survives by locking $U(1)_V$ with a suitable $SU(3)_{\rm strong}$ gauge transformation. If we denote the generator of $U(1)_V$ as $Q_V$, then the generator of the surviving global symmetry is defined as
\be Q_{\hat{V}} = Q_V + {\rm diag}\left(-1,-1,+2\right)_{\rm strong}\label{u1hat}\ee
At this point, the left-handed quarks $q_L$ and leptons $l_L$ have become gapped. We're left just with the right-handed quarks $q_R$, which now transform under the unbroken $SU(2)_{\rm strong}$ gauge group. Under the combined symmetry breaking %
\be SU(3)_{\rm strong} \times SU(2)_R \times U(1)_V\rightarrow SU(2)_{\rm strong}\times SU(2)_R \times U(1)_{\hat{V}}\nn\ee
the right-handed quarks decompose as 
\be q_R:  ({\bf 3},{\bf 2})_{+1} \rightarrow ({\bf 2},{\bf 2})_0 \oplus ({\bf 1},{\bf 2})_{+3}\label{qdecom}\ee
Now we let $SU(2)_{\rm strong}$ flow to the infra-red where it too confines. The $({\bf 2},{\bf 2})_0$ quarks above will form a condensate
\be \langle q_{R\,ai} q_{R\,bj}\rangle \sim \lstrong^3\epsilon_{abc}\hat{\sigma}^c \epsilon_{ij}\nn\ee
Here $\hat{\sigma}^c$ specifies the direction in $SU(3)$ colour space  determined by the weak condensate \eqn{colcond}. 
Importantly this new condensate breaks neither $SU(2)_R$ nor $U(1)_{\hat{V}}$. We're left with the infra-red symmetry which, once again, is unchanged in form from the ultra-violet \eqn{fuv},
\be F_{\rm weak} = SU(2)_R \times U(1)_{\hat{V}}\nn\ee
There is a single massless fermion transforming under $F_{\rm weak}$; this is the right-handed quark $\hat{q}_R$ that transforms in the $({\bf 1},{\bf 2})_{+3}$ representation in the decomposition \eqn{qdecom}, or
\FloatBarrier
\begin{table}[h!]
  \begin{center}
    \begin{tabular}{c||c:c}     
  &  $SU(2)_R$ & $U(1)_{\hat{V}}$ \\
     \hline
        $\hat{q}_R$  & ${\bf 2}$ & $+3$  \\
    \end{tabular}
  \end{center}
\end{table}
\FloatBarrier
\noindent
Once again this saturates the 't Hooft anomalies. Note, in particular, that the $U(1)$ charge $+3$ (as opposed to $-3$ seen when $\Lambda_{\rm strong}\gg \Lambda_{\rm weak}$) is compensated by the fact that we have a massless right-handed fermion, rather than left-handed. We see that the massless lepton $l_L$ in the regime $\Lambda_{\rm strong} \gg \Lambda_{\rm weak}$ has transmuted into a massless right-handed quark in the regime $\Lambda_{\rm strong}\ll \Lambda_{\rm weak}$. This provides a striking example of the lack of individual baryon and lepton number conservation in the theory. However, this transmutation occurs without violating the $B-L$ symmetry, a feat which is made possible by the twisting \eqn{u1hat} which means that the infra-red gauge-invariant down quark $\hat{q}_R$ carries a different $B-L$ quantum number from its gauge-dependent microscopic parent.

\subsubsection*{No Phase Transition?}

For a single generation, the global symmetry group of the theory remains unbroken both when $\lstrong\gg\lweak$ and when $\lweak\gg\lstrong$. While it is true that the UV symmetry group is locked with different gauge symmetries in each case, there is no gauge invariant way to distinguish them. This suggests that there is no phase transition as we vary the ratio $\lstrong/\lweak$, and the massless lepton transforms smoothly into the massless quark. 

This picture resonates with an old story. Recall that QCD with two flavours is one of the few cases where the 't Hooft anomalies can be saturated by massless baryons \cite{thooft}. There is a complementary way of viewing this \cite{tumbling,light,barnone}, in which a $\langle q_L q_L\rangle$ condensate forms,  Higgses $SU(3)_{\rm strong} \rightarrow SU(2)_{\rm strong}$, and leaving behind a massless quark. The fact that there is no phase transition between the Higgs and confining phases means that the massless baryon can be viewed as a continuously connected to this massless quark. 

Ultimately, the physics described in the above paragraph is thought not to occur for QCD. However, it does occur in the regime $\lweak\gg\lstrong$, where the $\langle q_Lq_L\rangle$ quark condensate \eqn{colcond} is encouraged by the $SU(2)_{\rm weak}$ force rather than  $SU(3)_{\rm strong}$. This suggests that as we head into the regime $\lstrong\approx \lweak$, it may be appropriate to better think of the massless quark $\hat{q}_R$ as a massless baryon. Indeed, the baryon $B\sim q_L\cdot q_L\cdot q_R$ has the same quantum numbers as the massless fermion. This means that, starting from the regime $\lstrong\gg \lweak$, the massless lepton can mix with the baryon, and ultimately emerge in the other regime $\Lambda_{\rm weak}\gg \Lambda_{\rm strong}$ as a massless quark.

\subsection{Multiple Generations}\label{nfmore}

We now repeat the analysis of the previous section, but with $N_f$ generations of fermions. The gauge group remains $G=SU(2)\times SU(3)$, but the global symmetry group is now (again omitting discrete factors)
\be F = SU(N_f)_{L'}\times SU(N_f)_L\times SU(2N_f)_R\times U(1)_V\label{f2}\ee
The quantum numbers of the fermions are
\FloatBarrier
\begin{table}[h!]
  \begin{center}
    \begin{tabular}{c||c:c|c:c:c:c} 
    & \multicolumn{2}{c|}{$G$}  &  \multicolumn{4}{c}{\ \ $F$}  \\
    
  & $SU(2)$  & $SU(3)$ & $SU(N_f)_{L'}$ & $SU(N_f)_L$ & $SU(2N_f)_R$ & $U(1)_V$ \\
     \hline
     $q_L$ & ${\bf 2}$ & ${\bf 3}$ & ${\bf 1}$ & ${\bf N}_f$ & ${\bf 1}$ & +1\\
      $l_L$ & ${\bf 2}$ & ${\bf 1}$ & ${\bf N}_f$ & ${\bf 1}$ & ${\bf 1}$ & $-3$  \\
       $q_R$ & ${\bf 1}$ & ${\bf 3}$ & ${\bf 1}$ & ${\bf 1}$ & ${\bf 2N}_f$ & $+1$ \\
    \end{tabular}
  \end{center}
\end{table}
\FloatBarrier
\noindent
There are now numerous 't Hooft anomalies for $F$. This time we will see that some of these symmetries are broken, with the 't Hooft anomalies in the surviving symmetries  saturated by massless fermions. 

Both $SU(2)_{\rm weak}$ and $SU(3)_{\rm strong}$ remain asymptotically free for $N_f\leq 5$. (This bound comes from the weak force; the strong force remains asymptotically free up to $N_f=8$ generations.) 

For suitably large $N_f$, the individual gauge theories sit in a conformal window while, for suitably low $N_f$, they undergo chiral symmetry breaking. The lower end of the conformal window is not well understood, but it is thought that it sits around 8 Dirac fermions for $SU(3)$ \cite{appel} and around 6 Dirac fermions for $SU(2)$ \cite{su21,su22,su23}. 

We analyse the theory in the regime in which both gauge groups undergo chiral symmetry breaking. This means that our analysis is restricted to $N_f=2$ and, possibly, $N_f=3$ which is a marginal case for $SU(2)_{\rm weak}$.

\subsubsection*{\underline{$\lstrong\gg \lweak$}}

Once again, the limit where the strong force dominates is well understood.  The usual QCD condensate forms, 
\be \langle q_{L\,i}^\dagger q_{R\,j}\rangle \sim \lstrong^3\Sigma_{ij}  \ \ \ \ i,j=1,2N_f\label{qcdcond}\ee
with a moduli space  parameterised by $\Sigma_{ij}$. If we ignore the weak force, then this condensate   breaks the $SU(2N_f)_L\times SU(2N_f)_R\rightarrow SU(2N_f)_{\rm diag}$ flavour symmetry in the usual fashion, resulting in $4N_f^2-1$ Goldstone bosons. 

We now turn on the $SU(2)_{\rm weak}$ gauge coupling. Often in such situations, different points on the moduli space give rise to different symmetry breaking patterns and one must work harder to determine which of the original possible vacua becomes the true vacuum \cite{peskin,preskill}. We will see a number of examples of this shortly. However, in the present situation this 
issue does not arise. Instead, each point in the moduli space breaks the $SU(2)_{\rm weak}$ gauge symmetry completely. 

The condensate \eqn{qcdcond} breaks the global symmetry group \eqn{f2} to 
\be F_{\rm strong} = SU(N_f)_{L'} \times SU(2)_{\rm diag} \times SU(N_f)_{\rm diag}\times U(1)_V\label{fstrong}\ee
where $SU(N_f)_{\rm diag} \subset SU(N_f)_L \times SU(2N_f)_R$ and, as in the previous section, $SU(2)_{\rm diag} \subset SU(2)_{\rm weak} \times SU(2N_f)_R$. This results in a moduli space of Goldstone modes,
\be {\cal M}_{\rm strong} = \frac{SU(N_f)_L \times SU(2N_f)_R}{SU(2) \times SU(N_f)_{\rm diag}}  \label{itsstrong}\ee
There are $\dim {\cal M}_{\rm strong} = 4(N_f^2-1)$ Goldstone bosons. Note that this is three fewer than the counting before we turned on $SU(2)_{\rm weak}$; these three were sacrificed on the altar of the Higgs mechanism. 

As in the previous section, the leptons remain massless. They transform under the surviving symmetry group $F_{\rm strong}$ as
\FloatBarrier
\begin{table}[h!]
  \begin{center}
    \begin{tabular}{c||c:c:c:c}     
  & $SU(N_f)_{L'}$&   $SU(2)_{\rm diag}$ & $SU(N_f)_{\rm diag}$ & $U(1)_V$ \\
     \hline
        $l_L$  & ${\bf N}_f$ & ${\bf 2}$ &  ${\bf 1}$ & $-3$  
    \end{tabular}
  \end{center}
\end{table}
\FloatBarrier
\noindent
It is simple to check that these  massless leptons saturate the 't Hooft anomalies of the surviving global symmetry $F_{\rm strong}$.

\subsubsection*{\underline{$\lweak\gg \lstrong$}}

When the weak force dominates, we again expect a condensate of left-handed fermions to form. We write the collection of $SU(2)_{\rm weak}$ doublets as $\psi_{mi}$, with $m=1,2,3,4$ labelling the quarks and leptons as in \eqn{psim}, and $i=1,\ldots,N_f$, the flavour index.
The condensate takes the general form
\be \langle \psi_{mi} \cdot \psi_{nj}\rangle\sim \lweak^3 \bar{J}_{mi,nj}\label{weakly}\ee
where we have suppressed the $SU(2)_{\rm weak}$ indices and $\bar{J}_{mi,nj}$ is a $4N_f\times 4N_f$ anti-symmetric  matrix.

If we ignore the strong force, then the $SU(2)_{\rm weak}$ gauge theory has   an $SU(4N_f)$ global symmetry which is broken by the condensate to $Sp(2N_f)$, resulting in an intermediate moduli space
\be {\cal M}_0  = \frac{SU(4N_f)}{Sp(2N_f)}\nn\ee
This is parameterised by $8N_f^2 - 2N_f -1$ Goldstone modes. The question that we want to answer is: what becomes of these modes when we turn on $SU(3)_{\rm strong}$?

This time there is a slightly involved calculation to do. Different choices of $\bar{J}_{mi,nj}$ give different symmetry breaking patterns for $SU(3)_{\rm strong}$ and a different mass spectrum for the resulting gauge bosons. For example, if the condensate forms in a flavour-diagonal fashion, with $\bar{J}_{mi,nj} = J_{mn}\delta_{ij}$, then it breaks the strong gauge group to 
\be SU(3)_{\rm strong} \rightarrow SU(2)_{\rm strong}\label{higgsing}\ee
which is the same symmetry breaking pattern that we saw in the $N_f=1$ case. Such a  condensate also breaks the global symmetry \eqn{f2} to 
\be \tilde{F} = SO(N_f) \times SU(2N_f)_R \times U(1)_{\hat{V}}\label{tildefisgood}\ee
where $SO(N_f) \subset SU(N_f)_{L'} \times SU(N_f)_L$ and $U(1)_{\hat{V}} \subset U(1)_V \times SU(3)_{\rm strong}$ as in \eqn{u1hat}.

Alternatively, a condensate $\bar{J}_{mi,nj}$ which is off-diagonal in the flavour basis will break the $SU(3)_{\rm strong}$ gauge group completely and  further break $\tilde{F}$ \cite{shrock}. (We provide a number of specific examples in Appendix \ref{unstablesec}.) The question that we must ask is: what is the preferred choice of breaking?

The tools to answer this question were introduced long ago by Peskin \cite{peskin} and Preskill \cite{preskill}. They showed how introducing a second gauge group induces a potential on the moduli space ${M}_0$. The true ground state of the system is determined by the minimum of this potential. We review this mechanism in some detail in Appendix \ref{asec}. Furthermore in Appendix \ref{SMalign} we show that the flavour-diagonal condensate, with symmetry breaking \eqn{higgsing} and \eqn{tildefisgood} is a local, stable minimum of the potential. Although we have been unable to prove, in generality, that there are not other local minima, we argue that generically one expects all other condensates to exhibit tachyonic modes and, in Appendix \ref{unstablesec}, we show this explicitly for a number of putative vacua with different symmetry breaking patterns.  The upshot is that the flavour-diagonal symmetry breaking pattern \eqn{higgsing} and \eqn{tildefisgood} appears to be dynamically preferred.

With the global symmetry $F$ defined in \eqn{f2} broken to $\tilde{F}$ in \eqn{tildefisgood}, the moduli space of ground states arising from the weak dynamics, is 
\be {\cal M}'_{\rm weak} = \frac{SU(N_f)_{L'}\times SU(N_f)_L}{SO(N_f)}\label{mprimeweak}\ee
Correspondingly, there are $\dim {\cal M}'_{\rm weak}  = \frac{3}{2}N_f^2 +\frac{1}{2}N_f -2$ Goldstone bosons. Note that, for $N_f>1$, the difference between $\dim {\cal M}_0$ and $\dim {\cal M}'_{\rm weak}$ is greater than the 5 Goldstone bosons eaten by the Higgs mechanism \eqn{higgsing}. This reflects the existence of a potential on  ${\cal M}_0$ induced by gauge symmetry $SU(3)_{\rm strong}$. We compute the masses of the resulting pseudo-Goldstone bosons in Appendix \ref{SMalign}.

We're still left with the dynamics of the unbroken $SU(2)_{\rm strong}$ gauge symmetry to contend with. Under the residual symmetry $SU(2)_{\rm strong}\times SU(2N_f)_R \times U(1)_{\hat{V}}$, the remaining quarks $q_R$ decompose as
\be q_R:  ({\bf 3},{\bf 2N}_f)_{+1} \rightarrow ({\bf 2},{\bf 2N}_f)_0 \oplus ({\bf 1},{\bf 2N}_f)_{+3}\label{qdecom2}\ee
When $SU(2)_{\rm strong}$ confines, the quarks in the $({\bf 2},{\bf 2N}_f)_0$ representation condense, further breaking the $SU(2N_f)_R$ global symmetry to $Sp(N_f)_R$. The final surviving global symmetry is
\be F_{\rm weak} = SO(N_f) \times Sp(N_f)_R \times U(1)_{\hat{V}}\label{fweak}\ee
and  Goldstone bosons parameterise the space
\be {\cal M}_{\rm weak} = \frac{SU(N_f)_{L'}\times SU(N_f)_L}{SO(N_f)} \times \frac{SU(2N_f)}{Sp(N_f)}\label{weakmod}\ee
As in the previous section, the massless fermion is now identified with $\hat{q}_R$, corresponding to the $({\bf 1},{\bf 2N}_f)_{+3}$ representation in \eqn{qdecom2}. This transformation properties of this fermion are %
\FloatBarrier
\begin{table}[!h]
  \begin{center}
    \begin{tabular}{c||c:c:c}     
  & $SO(N_f)$ &  $Sp(N_f)_R$ & $U(1)_{\hat{V}}$ \\
     \hline
        $\hat{q}_R$  & ${\bf 1}$ & ${\bf 2N}_f$ &  $+3$  
    \end{tabular}
  \end{center}
\end{table}
\FloatBarrier
\noindent
Again, the fermion $\hat{q}_R$ saturates the surviving 't Hooft anomalies.

In contrast to the case with $N_f=1$, the  symmetry breaking pattern \eqn{fstrong} and \eqn{fweak} differs in the two regimes, meaning that there is certainly a quantum phase transition as we vary the relative strengths of $\lstrong$ and $\lweak$. It is natural to ask the order of this phase transition.  

Sadly, the symmetry breaking structure gives little guidance. Note that neither  of the surviving symmetry groups, $F_{\rm strong}$ and $F_{\rm weak}$, is a subgroup of the other, reflecting the fact that the order parameters associated to the two phases are different. Most  phase transitions in Nature that exhibit this property are first order; indeed, this ``sub-group criterion" is  sometimes stated to be a clear indication of a first order phase transition. 
However, there is nothing that guarantees that this has to be the case. The two phases could be reached by two second order phase transitions, with an intermediate phase in between. This intermediate phase must have a global symmetry group that contains both $F_{\rm weak}$ and $F_{\rm strong}$ as subgroups, for example the UV global symmetry $F$. 

It is also possible that the transition proceeds through a single, continuous phase transition. In Landau theory, this requires  tuning to a multi-critical point. However, more exotic phase transitions, in which a gauge symmetry emerges and no fine tuning is needed, are also possible  \cite{deconfined}. Needless to say, it would be interesting to better understand the nature of the transition.

\subsection{Adding Hypercharge and Yukawa Couplings}\label{hypersec}

We now extend our study by including  $U(1)_Y$ hypercharge and Yukawa couplings. The gauge group is
\be G = U(1)_Y \times SU(2) \times SU(3)\nn\ee
To ensure cancellation of anomalies, we must now also include a right-handed electron $e_R$ in our theory.  We further include a single Higgs field, $\phi$. We will omit the right-handed neutrino for now, but revisit this in Section \ref{itsnewsec}. 

We include $N_f$ generations of fermions, coupled to the Higgs through the Yukawa couplings
\be {\cal L}_{\rm Yuk}  = \lambda_d\, q_{L\,i}^\dagger \phi\, d_{R\,i} + \lambda_u(q^\dagger_{L\,i}\cdot \phi^\dagger)\,u_{R\,i} + \lambda_e\,l_{L\,i}^\dagger \phi \,e_{R\,i}\label{yukawa}\ee
Here the flavour index $i=1,\ldots,N_f$ is summed over so that, in contrast to the Standard Model, there is an independent $SU(N_f)$ flavour symmetry for quarks and leptons, as well as the $B-L$ symmetry that we denote as $U(1)_V$
\be F = SU(N_f)_q \times SU(N_f)_l \times U(1)_V \label{fsm}\ee
The representations of the fields under $G$ and $F$ are shown in the table. 
\FloatBarrier
\begin{table}[h!]
  \begin{center}
    \begin{tabular}{c||c:c:c|c:c:c} 
    & \multicolumn{3}{c|}{$G$} & \multicolumn{3}{c}{$F$}    \\    
  & $U(1)_Y$ & $SU(2)$  & $SU(3)$  & $SU(N_f)_q$ & $SU(N_f)_l$ & $U(1)_V$   \\
     \hline
     $q_L$ & $+1$ &  ${\bf 2}$ & ${\bf 3}$ &${\bf N}_f$ &$ {\bf 1}$ & $+1$   \\
      $l_L$ & $-3$ & ${\bf 2}$ & ${\bf 1}$  & ${\bf 1}$ &$ {\bf N}_f$ & $-3$   \\
       $d_R$ & $-2$& ${\bf 1}$ & ${\bf 3}$  & ${\bf N}_f$ &$ {\bf 1}$ & $+1$  \\
       $u_R$ & $+4$ & ${\bf 1}$ & ${\bf 3}$ & ${\bf N}_f$ &$ {\bf 1}$ & $+1$  \\
        $e_R$ & $-6$ & ${\bf 1}$ & ${\bf 1}$ & ${\bf 1}$ &$ {\bf N}_f$ & $-3$  \\
        $\phi$ & $+3$ & ${\bf 2}$ & ${\bf 1}$ & ${\bf 1}$ &$ {\bf 1}$ & $0$ 
    \end{tabular}
  \end{center}
\end{table}
\FloatBarrier
%
%
  %
%
\noindent
Because we haven't included a right-handed neutrino, the $SU(N_f)_l\times U(1)_V$ symmetries have various 't Hooft anomalies, all of which arise from the leptons. The contribution to the 't Hooft anomalies from quarks vanish. 

We are interested in this theory in the regime 
\be v \ll \Lambda_{\rm weak},\Lambda_{\rm strong}\nn\ee
where the Higgs expectation value, $v$, is much smaller than all other scales so that the dynamics is dominated by the gauge interactions. 
We now repeat the analysis of previous sections. As before, we assume that $N_f$ is sufficiently small so that both gauge groups undergo chiral symmetry breaking; $N_f=2$ appears to surely be safe; $N_f=3$ is unclear.

\subsubsection*{\underline{$\lstrong\gg \lweak$}}

When the strong force dominates, a condensate \eqn{qcdcond} forms as before.   In terms of the up and down quarks, this reads
\be \langle q^\dagger_{L1i} d_{Rj}\rangle \sim \lstrong^3\delta_{ij} \ \ \ {\rm and}\ \ \ \langle q^\dagger_{L2i} u_{Rj}\rangle \sim \lstrong^3\delta_{ij} \nn\ee
where the $1,2$ labels on $q_L$ are $SU(2)_{\rm weak}$ indices. As in the Standard Model, this condensate breaks
\be U(1)_Y \times SU(2)_{\rm weak} \rightarrow U(1)_Q\nn\ee
where the generator of $U(1)_Q$ is related to the generator of $U(1)_Y$ by
\be Q = \frac{1}{6}Y + \frac{1}{2}{\rm diag}(1,-1)_{\rm weak}\nn\ee
This, of course, is the usual symmetry breaking pattern of electroweak down to electromagnetism.

As the theory   no longer has a  chiral symmetry, the full global symmetry  \eqn{fsm} survives in the infra-red 
\be F_{\rm strong} = SU(N_f)_q\times SU(N_f)_l \times U(1)_V\nn\ee
Because the full symmetry group survives, there are no Goldstone bosons. There are, however, light pion modes. These are the usual massless Goldstone bosons arising from the chiral symmetry breaking of QCD, which get a mass through the Yukawa couplings. (Even in the absence of a Higgs vev, the pions get a mass through mixing with $\phi$.) Some aspects of these pions, as well as the associated baryons, were discussed in  \cite{quigg}.

The leptons $l_L$ and $e_R$ remain unaffected by the gauge dynamics. They are distinguished by their charges under $U(1)_Q$; the left-handed lepton splits into $e_L$ with charge $Q=-1$ and $\nu_L$ with charge $Q=0$. Meanwhile, the right-handed electron $e_R$ has charge $Q=-1$. The electron pair gets a mass through the Yukawa coupling, while the left-handed neutrino remains massless, transforming as 
\FloatBarrier
\begin{table}[!h]
  \begin{center}
    \begin{tabular}{c||c|c:c:c}     
  & $U(1)_Q$   & $SU(N_f)_q$ &  $SU(N_f)_l$ & $U(1)_{V}$ \\
     \hline
        $\nu_L$  &0 &  ${\bf 1}$ & ${\bf N}_f$ &  $-3$  
    \end{tabular}
  \end{center}
\end{table}
\FloatBarrier
\noindent
This saturates the 't Hooft anomalies of $F$.

\subsubsection*{\underline{$\lweak\gg \lstrong$}}

When the weak force dominates, the condensate \eqn{weakly} forms. When we subsequently turn on both $SU(3)_{\rm strong}$ and $U(1)_Y$ gauge groups, we must again determine the correct vacuum.

One might be tempted to think that since $U(1)_Y$ is free in the infra-red, it does not affect the vacuum state described in the previous section. This, it turns out, is correct but it takes a calculation to show it. Indeed, in \cite{preskill}, Preskill gave examples of chiral symmetry breaking where a subsequent gauging of a $U(1)$ subgroup changes the vacuum structure when the $U(1)$ coupling constant becomes sufficiently strong. In  Appendix \ref{hyperalign} we show that this doesn't happen in the present case.

The upshot of this argument is that the condensate \eqn{weakly} that minimises the potential remains unchanged by $U(1)_Y$.  The quarks once again condense in a flavour-diagonal basis, as in \eqn{colcond}, to 
\be \langle q_{L\,ai}\cdot q_{L\,bj}\rangle \sim \lweak^3 \epsilon_{abc}\sigma^c \delta_{ij}\label{strongy}\ee
with $a,b,c=1,2,3$ colour indices and $i,j=1,\ldots N_f$ flavour indices. The remaining condensate pairs the $\hat{q}_{L\,i} = \sigma^aq_{L\,ai}$ quark with the leptons as in \eqn{colweakcond}
\be \langle \hat{q}_{Li} l_{Lj}\rangle \sim \lweak^3 \delta_{ij}\label{utrecht}\ee
The condensate breaks the global symmetry $F$ in \eqn{fsm} to 
\be F_{\rm weak} = SO(N_f)\times U(1)_{\hat{V}}\label{fweaker}\ee
where $SO(N_f) \subset SU(N_f)_q\times SU(N_f)_l$ and, as previously, $U(1)_{\hat{V}} \subset U(1)_V \times SU(3)_{\rm strong}$.

The two condensates break the remaining gauge group to 
\be SU(3)_{\rm strong} \times U(1)_Y \rightarrow SU(2)_{\rm strong} \times U(1)_{\hat{Q}}\label{hatqq}\ee
We have seen the  breaking to $SU(2)_{\rm strong}$ previously. To see that a $U(1)_{\hat{Q}}$ survives, note that 
both of the condensates \eqn{strongy} and \eqn{utrecht} carry $U(1)_Y$ charge $+2$. If we pick $\sigma^c=(0,0,1)$ in the condensate \eqn{strongy}, then we can construct the unbroken gauge generator
\be \hat{Q} = \frac{1}{6}Y - \frac{1}{6}{\rm diag}\left(1,1,-2\right)_{\rm strong}\nn\ee
The surviving $SU(2)_{\rm strong}$ gauge symmetry is coupled to the  
right-handed quarks. Under the breaking
\be SU(3)_{\rm strong} \times U(1)_Y \times U(1)_V \rightarrow SU(2)_{\rm strong}\times U(1)_{\hat{Q}} \times U(1)_{\hat{V}}\nn\ee
the right-handed fermions decompose as
\be d_R: &&{\bf 3}_{[-2,+1]} \rightarrow {\bf 2}_{[-\frac{1}{2},0]} \oplus {\bf 1}_{[0,+3]} \nn\\ 
u_R:&& {\bf 3}_{[+4,+1]} \rightarrow {\bf 2}_{[+\frac{1}{2},0]} \oplus {\bf 1}_{[1,+3]} \nn\\ 
e_R:&& {\bf 1}_{[-6,-3]} \rightarrow {\bf 1}_{[-1,-3]} 
\nn\ee
The fermions that transform as doublets under  $SU(2)_{\rm strong}$ condense and become gapped as the gauge group becomes strong. The resulting condensate does not further break $F_{\rm weak}$ from \eqn{fweaker}. This means that, in contrast to the regime $\lstrong\gg \lweak$,  there is now a moduli space of Goldstone
\be {\cal M}_{\rm weak} = \frac{SU(N_f)_q\times SU(N_f)_l}{SO(N_f)}\nn\ee
We're left with three gapless Weyl fermions, which were singlets under $SU(2)_{\rm strong}$.  Two of these, arising from $u_R$ and $e_R$, carry  equal and opposite $U(1)_{\hat{Q}}\times U(1)_{\hat{V}}$ charge. Although these are not coupled directly through the Yukawa coupling \eqn{yukawa}, there is nothing to prohibit this  pair becoming gapped as they interact with the scalar field. This leaves $\hat{d}_R$, the neutral component of the down quark, as the surviving massless fermion. It transforms as
\FloatBarrier
\begin{table}[!h]
  \begin{center}
    \begin{tabular}{c||c|c:c}     
  & $U(1)_{\hat{Q}}$   & $SO(N_f)$ &  $U(1)_{\hat{V}}$ \\
     \hline
        $\hat{d}_R$  &0 &  ${\bf N}_f$ &  $+3$  \\
    \end{tabular}
  \end{center}
\end{table}
\FloatBarrier
\noindent
It is noticeable that in the UV theory, the quarks did not appear to play any role in the computation of 't Hooft anomalies. Yet, by the time we flow to the infra-red, the sole remaining fermion is a quark and saturates the surviving 't Hooft anomalies of $F_{\rm weak}$. 

Note that in both $\lstrong\gg \lweak$ and $\lweak \gg \lstrong$ regimes, there is a surviving $U(1)$ gauge symmetry that we may identify with electromagnetism, and a surviving $U(1)$ global symmetry that we may identify with $B-L$. These symmetries are twisted with different gauge symmetries in the two regimes, but this does not impede us from identifying them. This conclusion differs from \cite{quigg} were it is claimed that both electromagnetic and $B-L$ symmetries are broken in the $\lweak\gg \lstrong$ regime. 

The addition of hypercharge and Yukawa couplings does not change the conclusions of our earlier sections. If we have $N_f=1$ generation of fermions, then it seems plausible that the transition between the two regimes proceeds without a phase transition. Meanwhile, for $N_f\geq 2$, a phase transition must occur. 

However, in contrast to the situation in Section \ref{nfmore}, there is a fairly simple symmetry breaking pattern between the two regimes, with $SU(N_f)_q\times SU(N_f)_l$, which survives when the strong force dominates, breaking to $SO(N_f)_{\rm diag}$ when the weak force dominates, suggesting that a mean-field description of the phase transition in terms of Landau theory may be appropriate. 


\section{A Novel Chiral Gauge Theory}\label{itsnewsec}

In this section, we extend our analysis to a chiral gauge theory with gauge group
\be G = U(1)_Y \times Sp(r) \times SU(N)\nn\ee
Anomaly considerations, to be described below, mean that we must take $N$ odd. For the simplest values of $r=1$ and $N=3$ this gauge group coincides with that of the  Standard Model.

The chiral fermion content is a natural extension of that of the Standard Model: we take left-handed fermions $q_L$ and $l_L$ to  transform under $Sp(r)$, while the right-handed fermions are singlets under $Sp(r)$. One key difference is that we must take $r$ copies of each of the right-handed fermions, including $r$ copies of the right-handed neutrinos $\nu_R$. We introduce an index $\alpha=1,\ldots ,r$ to distinguish these fields. For later convenience, we also introduce $r$ distinct Higgs fields $\phi_\alpha$ at this time too. The full set of fermions and scalars and their transformations is given by
\FloatBarrier
\begin{table}[h!]
  \begin{center}
    \begin{tabular}{c||c:c:c} 
 
  & $U(1)_Y$ & $Sp(r)$  & $SU(N)$   \\
     \hline
     $q_L$ & $+1$ &  ${\bf 2r}$ & ${\bf N}$    \\
      $l_L$ & $-N$ & ${\bf 2r}$ & ${\bf 1}$    \\
       $d_{R\alpha}$ & $-(2\alpha-1)N+1$& ${\bf 1}$ & ${\bf N}$    \\
       $u_{R\alpha}$ & $+(2\alpha-1)N+1$ & ${\bf 1}$ & ${\bf N}$   \\
        $e_{R\alpha}$ & $-2\alpha N$ & ${\bf 1}$ & ${\bf 1}$   \\
         $\nu_{R\alpha}$ & $(2\alpha-2)N$ & ${\bf 1}$ & ${\bf 1}$   \\
        $\phi_\alpha$ & $(2\alpha-1)N$ & ${\bf 2r}$ & ${\bf 1}$ 
    \end{tabular}
  \end{center}
\end{table}
\FloatBarrier
\noindent
with $\alpha =1,\ldots,r$. 
It is straightforward to show that, with these charge assignments,  all gauge anomalies vanish. The ${\bf Z}_2$ anomaly of $Sp(r)$ is vanishing only for $N$ odd. Notice that the first right-handed neutrino is decoupled from the gauge fields, as in the Standard Model, but the other $r-1$  carry $U(1)_Y$ charge. 

The $U(1)_Y$ charge assignments also allow us to construct Yukawa interactions. For a single generation, we have
\be {\cal L}_{\rm Yuk}  = \lambda_d\,q_L^\dagger \phi_\alpha d_{R\alpha} + \lambda_u(q_L^\dagger \cdot\phi^\dagger_{\alpha})u_{R\alpha} + \lambda_e\,l_L^\dagger \phi_\alpha e_{R\alpha} + \lambda_\nu(l_L^\dagger \cdot \phi^\dagger_\alpha) \nu_{R\alpha} \label{yukky}\ee
Here $(q_L^\dagger\cdot\phi^\dagger)$ denotes the $Sp(r)$ singlet that one can construct from these two fields. (This is analogous to the $SU(2)$ singlet constructed using $\epsilon_{ab}$.)

The generalisation of the Standard Model with $SU(3)$ gauge group replaced by $SU(N)$ is fairly well explored. (See, for example, \cite{more}.) The generalisation with $SU(2)$ replaced by $Sp(r)$, with the anomaly-free charge assignments shown in the table above, appears to be novel. We note the possibility that such a theory with gauge group  $U(1)\times Sp(r) \times SU(3)$ may describe our world, with the additional Higgs fields $\phi_\alpha$, $\alpha =2,\ldots,r$, breaking $Sp(r)$ to $SU(2)$ at some high scale. Moreover, this  two parameter extension of the Standard Model may lend itself to a large $r$, large $N$ expansion; we leave this possibility to future work. (A large $N$ expansion of certain chiral gauge theories was previously proposed in \cite{eichten,adi}.)

We will adopt the convention of the Standard Model and refer to the $Sp(r)$ gauge group as {\it weak} and the $SU(N)$ gauge group as {\it strong}. As in the previous section, we will be interested in the phase diagram of the theory, with the two asymptotic regimes in which one of the  gauge groups dominates over the other. We will study different variants of this problem, both with and without hypercharge interactions and Yukawa couplings.

\subsubsection*{Beta Functions}

We will discuss the chiral theory coupled to $N_f$ generations of fermions and focus on situations where both gauge groups are asymptotically free . The $SU(N)$ gauge group is coupled to $2rN_f$ Dirac fermions, each in the fundamental representation, and is asymptotically free provided
\be 11N > 4rN_f\nn\ee
%
%
%
Meanwhile, the $Sp(r)$ factor is coupled to $N_f(N+1)$ Weyl fermions, each in the pseudo-real fundamental representation. If we ignore the Higgs fields for now, $Sp(r)$ is asymptotically free provided
\be 11(r+1)  > (N+1) N_f\nn\ee
%
%
For $N_f\geq 6$, at least one of the gauge groups is infra-red free. In contrast, for any  $N_f\leq 5$, there are always choices of $N$ and $r$ for which both gauge groups  become strongly coupled in the infra-red. This conclusion persists in the presence of Higgs fields.

%
%

As before, our analysis will rely on chiral symmetry breaking in the regime where one or the other gauge group becomes strongly coupled. This takes place for suitably low $N_f$, below the conformal window. The lower-edge of the conformal window is  not well established. The $SU(N)$ gauge factor has $2rN_f$ Dirac fermions in the fundamental representation, and undergoes chiral symmetry breaking for
\be 2r N_f < C_\star N\nn\ee
for some $C_\star$ which is expected to sit somewhere around 3 to 4. Meanwhile the $Sp(r)$ gauge factor has $N_f(N+1)$ Weyl fermions in the pseudo-real fundamental representation, and is expected to undergo chiral symmetry breaking when
\be N_f(N+1) < \hat{C}_{\star} (r+1)\nn\ee
where $\hat{C}_\star$ is around 6 to 8. (See, for example, \cite{frandsen}.) For  $N_f\leq 2$ there are an infinite number of choices of  $N$ and $r$ for which chiral symmetry breaking occurs, while for $N_f=4$ it seems likely there are none. The situation for $N_f=3$ is, in all cases, more murky.

\subsection{$Sp(r) \times SU(N)$}

We start by neglecting the $U(1)_Y$ factor and focussing only on the gauge group
\be G = Sp(r) \times SU(N)\nn\ee
Because the right-handed electrons $e_R$ and neutrinos $\nu_R$ are singlets under the non-Abelian part of the gauge group, we may ignore them for the purpose of this discussion. We will also discard the Higgs field for now, focussing only on the fermions. As in the case of the Standard Model, here we will find the richest symmetry breaking patterns, unconstrained by hypercharge assignments and Yukawa couplings. We will then reintroduce both of these in Section \ref{andagain}

With $N_f$ generations of fermions, the global, non-anomalous, symmetry group is
\be F = SU(N_f)_{L'} \times SU(N_f)_L \times SU(2rN_f)_R \times U(1)_V\label{fspr}\ee
Under the gauge and global symmetry groups, the fermions transform as
\FloatBarrier
\begin{table}[h!]
  \begin{center}
    \begin{tabular}{c||c:c|c:c:c:c} 
    & \multicolumn{2}{c|}{$G$}  &  \multicolumn{4}{c}{\ \ $F$}  \\
    
  & $Sp(r)$  & $SU(N)$ & $SU(N_f)_{L'}$ & $SU(N_f)_L$ & $SU(2rN_f)_R$ & $U(1)_V$ \\
     \hline
     $q_L$ & ${\bf 2r}$ & ${\bf N}$ & ${\bf 1}$ & ${\bf N}_f$ & ${\bf 1}$ & +1\\
      $l_L$ & ${\bf 2r}$ & ${\bf 1}$ & ${\bf N}_f$ & ${\bf 1}$ & ${\bf 1}$ & $-N$  \\
       $q_R$ & ${\bf 1}$ & ${\bf N}$ & ${\bf 1}$ & ${\bf 1}$ & ${\bf 2rN}_f$ & $+1$ 
    \end{tabular}
  \end{center}
\end{table}
\FloatBarrier
\noindent
with $q_R = (u_R,d_R)$, the two right-handed quarks now undistinguished by hypercharge.  There are numerous 't Hooft anomalies between the various subgroups of $F$.

\subsubsection*{\underline{$\lstrong\gg \lweak$}}

When the $SU(N)_{\rm strong}$ force dominates, the usual quark condensate \eqn{qcdcond} forms and the quarks become gapped, leaving behind a number Goldstone bosons. 

If we ignore the $Sp(r)$ weak force, the theory has a $SU(2rN_f)_L\times SU(2rN_f)_R \times U(1)_V$ global symmetry, broken by the condensate to $SU(2rN_f)_{\rm diag}\times U(1)_V$. Gauging $Sp(r)_{\rm weak}$, means that the global symmetry $F$ in \eqn{fspr} breaks to
\be F_{\rm strong} = SU(N_f)_{L'} \times Sp(r)_{\rm diag} \times SU(N_f)_{\rm diag}\times U(1)_V\nn\ee
Here $Sp(r)_{\rm diag} \subset Sp(r)_{\rm weak}\times SU(2rN_f)_{\rm diag}$ arises from a simultaneous gauge transformation and surviving $SU(2rN_f)_{\rm diag}$ rotation. This ensures that the $Sp(r)$ gauge symmetry is fully broken, with only this ``weak-flavour-locked" global symmetry surviving. The remaining $SU(N_f)_{\rm diag}$ is the centraliser of $Sp(r)$ in $SU(2rN_f)_{\rm diag}$. 

The Goldstone bosons therefore parameterise the moduli space
\be {\cal M}_{\rm strong} = \frac{SU(N_f)_L\times SU(2rN_f)_R}{Sp(r) \times SU(N_f)_{\rm diag}}\nn\ee
There are $\dim {\cal M}_{\rm strong} = 4r^2 N_f^2 - 2r^2 - r -1$ of them.

With the $Sp(r)_{\rm weak}$ gauge group fully Higgsed, the left-handed leptons remain massless. They transform under $F_{\rm strong}$ as
\FloatBarrier
\begin{table}[h!]
  \begin{center}
    \begin{tabular}{c||c:c:c:c}     
  & $SU(N_f)_{L'}$&   $Sp(r)_{\rm diag}$ & $SU(N_f)_{\rm diag}$ & $U(1)_V$ \\
     \hline
        $l_L$  & ${\bf N}_f$ & ${\bf 2r}$ &  ${\bf 1}$ & $-N$  \\
    \end{tabular}
  \end{center}
\end{table}
\noindent
These saturate the 't Hooft anomaly of $F_{\rm strong}$. 

\subsubsection*{\underline{$\lweak\gg \lstrong$}}

When the $Sp(r)_{\rm weak}$ force dominates, the $(N+1)N_f$ left-handed fermions condense. Collectively, we refer to these as $\psi_{mi}$, with $m=1,\ldots,(N+1)$ labelling the quarks and leptons in a single generation, and $i=1,\ldots,N_f$ the flavour index. The condensate takes the form
\be \langle \psi_{mi} \cdot \psi_{nj}\rangle\sim \lweak^3 \bar{J}_{mi,nj}\label{weaklydoesit}\ee
with $\psi_{mi} \cdot \psi_{nj}$ a $Sp(r)_{\rm weak}$ singlet. (The gauge group indices are contracted using the $Sp(r)$ invariant anti-symmetric tensor) and $\bar{J}_{mi,nj}$ an anti-symmetric matrix.

Once again, we must determine the choice of $\bar{J}_{mi,nj}$ that minimizes the potential induced by gauging the $SU(N)_{\rm strong}$ group. This is a fairly involved calculation and is presented in Appendix \ref{hardalign}, where we show that the flavour-diagonal condensate is again a (local) minimum of the potential, with no tachyonic modes. This means that the dynamically preferred vacuum condensate breaks  the gauge group to
\be SU(N)_{\rm strong} \rightarrow Sp((N-1)/2)_{\rm strong}\label{notthatstrong}\ee
generalising the earlier result \eqn{higgsing}. At the same time, the global symmetry $F$ is broken to 
\be \tilde{F} = SO(N_f)\times SU(2rN_f)_R\times U(1)_{\hat{V}}\label{formulastrong}\ee
where $SO(N_f)\subset SU(N_f)_{L'}\times SU(N_f)_L$ and $U(1)_{\hat{V}} \subset U(1)_V\times SU(N)_{\rm strong}$ is defined in analogy with \eqn{u1hat}. 

We're still left with the right-handed quarks $q_R$, which are now coupled to the surviving $Sp(\frac{1}{2}(N+1))$ gauge group. Under the symmetry breaking
\be SU(N)_{\rm strong} \times U(1)_V \rightarrow Sp((N-1)/2)_{\rm strong} \times U(1)_{\hat{V}}\nn\ee
these right-handed quarks decompose as
\be q_R: {\bf N}_{+1} \rightarrow ({\bf N-1})_0 \oplus {\bf 1}_{+N}\label{soju}\ee
As $Sp(\frac{1}{2}(N-1)$ becomes strong and confines, those quarks transforming in the ${\bf N-1}$ representation condense. This further breaks the flavour symmetry group to 
\be F_{\rm weak} = SO(N_f) \times Sp(rN_f)_R \times U(1)_{\hat{V}}\label{formulaweak}\ee
The final result is that we have a moduli space of vacua,
\be {\cal M}_{\rm weak} = \frac{SU(N_f)\times SU(N_f)}{SO(N_f)} \times \frac{SU(2rN_f)}{Sp(rN_f)}\nn\ee
generalising our earlier result \eqn{weakmod}. Meanwhile, the singlet fermions in \eqn{soju} remain massless, transforming under $F_{\rm weak}$ as
\FloatBarrier
\begin{table}[!h]
  \begin{center}
    \begin{tabular}{c||c:c:c}     
  & $SO(N_f)$ &  $Sp(rN_f)_R$ & $U(1)_{\hat{V}}$ \\
     \hline
        $\hat{q}_R$  & ${\bf 1}$ & ${\bf 2rN}_f$ &  $+N$  \\
    \end{tabular}
  \end{center}
\end{table}
\FloatBarrier
\noindent
For a single generation of fermions, $N_f=1$, we again see that the infra-red global symmetry in the two regimes coincides: both are $Sp(r)\times U(1)$. This now differs from the UV symmetry \eqn{fspr}, meaning that there are Goldstone bosons even in this case. Nonetheless, it appears plausible that there is no phase transition for $N_f=1$ as we vary the gauge coupling constants. As in Section \ref{nonabsec}, the left-handed lepton in one regime transmutes into a right-handed quark in the other.

For $N_f\geq 2$, the symmetry breaking patterns on either side differ and there must be a phase transition as we vary between them. Once again, neither  of the symmetries $F_{\rm strong}$ and $F_{\rm weak}$, defined in \eqn{formulastrong} and \eqn{formulaweak}, are subgroups  of the other.

\subsection{Adding Hypercharge and Yukawa Couplings}\label{andagain}

Finally we discuss the theory introduced at the beginning of this section, replete with $U(1)_Y$ coupling and flavour-diagonal Yukawa interactions \eqn{yukky}. The gauge symmetry is
\be G = U(1)_Y \times Sp(r) \times SU(N)\nn\ee
and the global symmetry is
\be F = SU(N_f)_q \times SU(N_f)_l \times U(1)_V\label{ffinal}\ee
where the matter fields transform as
\FloatBarrier
\begin{table}[h!]
  \begin{center}
    \begin{tabular}{c||c:c:c|c:c:c} 
    & \multicolumn{3}{c|}{$G$} & \multicolumn{3}{c}{$F$}    \\    
  & $U(1)_Y$ & $Sp(r)$  & $SU(N)$  & $SU(N_f)_q$ & $SU(N_f)_l$ &  $U(1)_V$   \\
     \hline
     $q_L$ & $+1$ &  ${\bf 2}$ & ${\bf 3}$ &${\bf N}_f$ &$ {\bf 1}$ &  $+1$   \\
      $l_L$ & $-N$ & ${\bf 2}$ & ${\bf 1}$  & ${\bf 1}$ &$ {\bf N}_f$ &  $-N$   \\
       $d_{R\alpha}$ & $-(2\alpha-1)N+1$   & ${\bf 1}$ & ${\bf 3}$  & ${\bf N}_f$ &$ {\bf 1}$ &  $+1$  \\
       $u_{R\alpha}$ & $+(2\alpha-1)N+1$  & ${\bf 1}$ & ${\bf 3}$ & ${\bf N}_f$ &$ {\bf 1}$ &  $+1$  \\
        $e_{R\alpha}$ & $-2\alpha N$ & ${\bf 1}$ & ${\bf 1}$ & ${\bf 1}$ &$ {\bf N}_f$  &  $-N$  \\
        $\nu_{R\alpha}$ & $(2\alpha-2)N$ & ${\bf 1}$ & ${\bf 1}$   & ${\bf 1}$ &$ {\bf N}_f$  &  $-N$  \\
       $\phi_\alpha$ & $(2\alpha-1)N$ & ${\bf 2}$ & ${\bf 1}$ & ${\bf 1}$ &$ {\bf 1}$&  $0$ 
    \end{tabular}
  \end{center}
\end{table}
\FloatBarrier
\noindent
This time, the presence of the right-handed neutrinos ensure that the global symmetries $F$ suffer no 't Hooft anomalies.
The arguments of the previous section allow us to quickly determine the symmetry breaking pattern in the two regimes. 

\subsubsection*{\underline{$\lstrong\gg \lweak$}}

When the strong force dominates, the full UV symmetry \eqn{ffinal} survives. There are no Goldstone bosons. 
%
%
Although the leptons remain massless after the gauge interactions become strong, they interact with the Higgs fields and, indirectly, with the mesons and there is nothing to prevent them gaining a mass, suppressed by the Yukawa coupling. For generic values of the Yukawa couplings, we therefore expect the fermions to be gapped.

\subsubsection*{\underline{$\lweak\gg \lstrong$}}

When the weak force dominates, the condensate \eqn{weakly} forms. A computation of the correct vacuum alignment can be found in Appendix \ref{hardalign}, which shows that the ground state preserves the global symmetry
\be  {F}_{\rm weak} = SO(N_f) \times U(1)_{\hat{V}}\nn\ee
with $SO(N_f) \subset SU(N_f)_{\rm diag} \subset SU(N_f)_q\times SU(N_f)_l$.  The condensate also breaks the $SU(N)_{\rm strong}$ gauge group as in \eqn{notthatstrong}. As the surviving subgroup of $SU(N)_{\rm strong}$ confines, the resulting condensate  does not further break the global symmetry $F_{\rm weak}$. 
Once again, no symmetry principle ensures massless fermions and, generically, none are expected to survive. Instead, the gapless modes are supplied by the Goldstone bosons which parameterise
\be {\cal M}_{\rm weak} = \frac{SU(N_f)_q\times SU(N_f)_l}{SO(N_f)}\label{onelastm}\ee
The story is, by now, familiar. There is no evidence of a phase transition in the symmetry breaking pattern when $N_f=1$. Such a phase transition must occur for $N_f\geq 2$ although now the symmetry breaking pattern suggests that such a phase transition can be captured by a mean field Landau-Ginzburg description.  It would surely be interesting to gain a better understanding of the nature of the phase transition, both here and in other examples. 
 
\section*{Acknowledgements}

 DT is grateful to KIAS for  their kind hospitality while this work was done. 
We are supported  by the STFC consolidated grant ST/P000681/1. DT is a Wolfson Royal Society Research Merit
Award holder and is supported by a Simons Investigator Award. NL is supported by the Internal Graduate Scholarship from Trinity College, Cambridge, and by the Thai Government.


\setcounter{section}{0}
\renewcommand\thesection{\Alph{section}}
\renewcommand\thesubsection{\thesection.\arabic{subsection}}

\section{Appendix: Vacuum Alignment}\label{asec}

When a gauge theory spontaneously breaks chiral symmetry, the resulting Goldstone bosons parameterise a moduli space of vacua ${\cal M}_0$. If this theory is subsequently coupled to a second gauge group, which becomes strong at a lower scale, then much of the the vacuum moduli space ${\cal M}_0$ is lifted. The preferred ground state is chosen dynamically in a process known as {\it vacuum alignment}. 

The physics of vacuum alignment was explained in two beautiful papers by Peskin \cite{peskin} and Preskill \cite{preskill}.
In this appendix, we review the results of these papers. In Appendix \ref{bsec}, we then   apply these results to understand the ground states in situations of interest in Sections \ref{nonabsec} and \ref{itsnewsec}.

We consider a general gauge theory with gauge group as 
\be G = G_1\times G_2\nn\ee
with the convention that the gauge group $G_1$ will always run to strong coupling before $G_2$, meaning that the dynamically generated scales are ordered as
\be \Lambda_1 \gg \Lambda_2\nn\ee
%
We couple our gauge theory to fermions. 
The full theory will have a global symmetry group that we denote as $F$. However, if we first turn off the second gauge group $G_2$ by setting its coupling to zero, the global symmetry group of remaining theory, with only $G_1$, will  be larger: we denote this global symmetry group as $K$. 

We are interested in situation where the confinement of $G_1$ and subsequent condensation of fermion bilinears breaks this global symmetry to a smaller subgroup
\be K \longrightarrow H \nn\ee
The are three basic symmetry breaking patterns, suggested by the maximally attractive channel hypothesis,  that can arise with a single gauge group $G_1$ \cite{peskin}. Only two of them will be needed in the bulk of the paper, but we list all three for completeness:
\begin{itemize}
\item If there are $n$ massless Dirac fermions in a complex representation of $G_1$, we have the global symmetry $K=SU(n)_L\times SU(n)_R$, with the two factors acting on left- and right-handed Weyl fermions which we denote as $\psi_L$ and $\psi_R$.  The condensate takes the general form
\be \langle \psi_{R\,i}^\dagger\psi_{L\,j}\rangle \sim \Lambda_1^3\, (U^\dagger_R U_L)_{ij}\nn\ee
with $i,j=1,\ldots,n$ and $U_{L/R}\in SU(N)_{L/R}$.  The subgroup $H=SU(n)_{\rm diag}$ leaves the condensate untouched, meaning that we have  the familiar QCD-like breaking pattern 
\be SU(n)_L\times SU(n)_R \longrightarrow SU(n)_{\diag} \nn\ee
This form of condensate arises in the bulk of the paper when the $SU(N)$ gauge group, with $N\geq 3$, first becomes strong. 

\item
 If there are $2n$ Weyl fermions in a pseudo-real representation of $G_1$ then we have a global symmetry $K=SU(2n)$. The condensate forms through the invariant anti-symmetric tensor, $\epsilon^{ab}$. It takes the general form 
\be \left<\epsilon^{ab}\psi_{a\,i}\psi_{b\,j}\right> \sim \Lambda_1^3\ (U^TJU)_{ij}\ \ \ \ \ \ \ {\rm with}\ J=\left(\begin{array}{cc} 0 & 1 \\ -1 & 0 \end{array}\right)\nn\ee
Here $U\in SU(2n)$ and  $J$ is the  $2n\times 2n$ anti-symmetric matrix given in block form above. The resulting symmetry breaking pattern is 
\be SU(2n) \longrightarrow Sp(n) \nn\ee
This form of the condensate arises in the bulk of the paper when the gauge group $Sp(r)$ or $SU(2)$ becomes strong. 
\item
If there are $2n$ Weyl fermions in a real representation of $G_1$, then the global symmetry is again $K=SU(2n)$. This time the condensate forms through the invariant symmetry tensor of the reperesentation, $d^{ab}$. It takes the general form
\be \left<d^{ab} \psi_{a\,i}\psi_{b\,j}\right> \sim \Lambda_1^3\,(U^TDU)_{ij}  \ \ \ \ \ \ {\rm with}\ D=\left(\begin{array}{cc} 0 & 1 \\ 1 & 0 \end{array}\right)\nn\ee
Again,  $U\in SU(2n)$ and  $D$ is a  $2n\times 2n$ symmetric matrix given in block form above. Now, the symmetry breaking pattern is 
\be SU(2n) \longrightarrow O(2n)\nn\ee
This form of the condensate does not play a direct role in this paper although, as we show in Section \ref{nfmore},  we have a similar symmetry breaking pattern when first $Sp(r)$ and subsequently $SU(N)$ becomes strong. 
\end{itemize}
Each of the symmetry breaking patterns described above results in a vacuum moduli space
\be {\cal M}_0 = K/H\nn\ee
Each point of ${\cal  M}_0$ corresponds to a different  orientation of $H\subset K$.  

\subsection{A Potential Over the Moduli Space}

We now turn on the coupling for the second gauge group $G_2\subset K$. The global symmetry of the theory is  reduced to $F$. Correspondingly, the symmetry breaking pattern $K\rightarrow H$  is reduced.  Different orientations of $H$ in $K$ descend to different symmetry breaking patterns, each of the form
\be G_2\times F \longrightarrow \tilde{G} \times \tilde{F}\label{survive}\ee
The question we need to address is: what symmetry breaking pattern is preferred? This is the question of {\it vacuum alignment}.

As explained in \cite{peskin,preskill}, the choice of vacuum is determined dynamically. To see why this is the case note that, after gauging $G_2$, there are three different fates for the  would-be Goldstone modes in ${\cal M}_0$. Some will be charged under $G_2$; these act as Higgs bosons, breaking $G_2$ to the smaller group $\tilde{G}\subset G_2$ and  are eaten by the Higgs mechanism.  Other scalars in ${\cal M}_0$ are not eaten, but are no longer protected by symmetry constraints;  they will gain a mass, as we explain more fully below, and  are referred to as {\it pseudo-Goldstone bosons}. Finally, some scalars remain exactly massless; these are  Goldstone modes of the full theory, whose moduli space includes the factor
\be {\cal M}\subset F/\tilde{F}\nn\ee
Note that this need not be the full moduli space because when the surviving gauge group $\tilde{G}$ becomes strong, it too may break some chiral symmetry, resulting in further Goldstone bosons. 

We now describe how the potential is generated over ${\cal M}_0$, following \cite{peskin,preskill}. The minimum determines the locus of ground states in  ${\cal M}_0$ and, correspondingly, the surviving symmetries $\tilde{G}$ and $\tilde{F}$ in \eqn{survive}. The potential is generated by the coupling
\be \delta {\cal L} = \sum_\alpha A_\mu^\alpha J^{\alpha\mu}\nn\ee
with $\alpha=1,\ldots,\dim G_2$. The current is given by
\be J^\alpha_\mu = i\bar{\psi}\gamma_\mu G^\alpha \psi\nn\ee
where $G^\alpha$ are the generators of $G_2$.

At one-loop, exchange of the W-bosons gives rise to a potential on the moduli space. A point $\Omega\in {\cal M}_0$ corresponds to a putative vacuum state $\ket{\Omega}$. The energy of this state is given by
\be
V(\Omega) = -\frac{g_2^2}{2} \mes \Delta^{\mu\nu}(x)\bra{\Omega}{J}^{\alpha}_{\mu}(x){J}^{\alpha}_{\nu}(0)\ket{\Omega}
\label{vomega}\ee
This correlation function, and those below, are  time-ordered. Here $g_2$ is the gauge coupling associated to the gauge group $G_2$ and the  gauge propagator $\Delta_{\mu\nu}(x)$ is defined in the usual way by
\be
\bra{\Omega} A^{\alpha}_{\mu}(x) A^{\beta}_\nu (0)\ket{\Omega} = -i \delta^{\alpha \beta} \Delta_{\mu\nu}(x)\nn
\ee
It will prove to be useful to change perspective, somewhat analogous to the shift from an active to passive viewpoint. To this end, we fix a reference vacuum state $\ket{0}$. A general ground state $\ket{\Omega}$ is given by the unitary action
\be \ket{\Omega} = U\ket{0}\nn\ee
with $U \in K/H$. (Strictly speaking, $U$ is a unitary representation of $K$ acting on the Hilbert space.) We now parameterise the point in ${\cal M}_0$ by $U\in K/H$. We define the rotated currents, 
\be \mathcal{J}^\alpha_\mu= i\bar{\psi}\gamma_\mu U^\dagger G^\alpha U \psi\label{ucurrent}\ee
In this notation, the potential \eqn{vomega} becomes
\be
V(U) = -\frac{g_2^2}{2}\mes\Delta^{\mu\nu}(x)\bra{0}\mathcal{J}^{\alpha}_{\mu}(x)\mathcal{J}^\alpha_{\nu}(0)\ket{0}\label{potj}
\ee
In the vacuum $\ket{0}$, there is a particular embedding of the unbroken subgroup $H\subset K$. We introduce the following notation for the generators of the Lie algebra of $K$ and its sub-algebras\footnote{To avoid an explosion of notation, we  denote the Lie group and Lie algebra by the same letter.}
\begin{itemize}
\item  Let $T^m$, $m=1,\ldots, \dim K$, denote the generators of $K$
\item Let  $H^a$, $a=1,\ldots,\dim H$, denote the generators of $H\subset K$.
\item Let $X^i$, $i=1,\ldots, \dim K - \dim H$, denote the generators of $K/H$. 
\end{itemize}
Any generator, $T$ of $K$, can be decomposed into components projected along the two sub-algebras $H$ and $X=K/H$. We write the projection along $H$ as $T_H$ and the projection along $X$ as $T_X$, so we have
\be T = T_H + T_X :=  {\rm Tr}(TH^a)\, H^a + {\rm Tr}(TX^i) X^i\nn\ee
with $T_H^a = {\rm Tr}(TH^a)$ the projection along $H$ and $T_X^i = {\rm Tr}(TX^i)$ the projection along $X$. Implementing a decomposition of this kind for  the current \eqn{ucurrent}, we have
\be \mathcal{J}^\alpha_\mu =  {\rm Tr}(U^\dagger G^\alpha UH^a)\, {\mathcal J}^a_{H\mu} +  {\rm Tr}(U^\dagger G^\alpha UX^i) {\mathcal{J}}^i_{X\mu}\nn\ee
with ${\mathcal J}^a_{H\mu} = i\bar{\psi}\gamma_\mu  H^a \psi$ the currents that lie in the unbroken $H\subset K$ and ${\mathcal J}^i_{X\mu} = i\bar{\psi}\gamma_\mu  X^i \psi$ the currents that lie in the broken $K/H$.  Substituting this decomposition into the potential \eqn{potj}, we have three terms: ${\cal J}_H^2$, ${\cal J}_X^2$ and ${\cal J}_H{\cal J}_X$. The cross-term ${\cal J}_H{\cal J}_X$ vanishes. The other two terms also simplify. In particular, using the fact that $K/H$ is a symmetric space, we have
\be
\bra{0}{\cal J}^i_{X\mu}(x) {\cal J}^{j}_{X\nu} (0)\ket{0} = \Tr(X^i X^j)\bra{0}{\cal J}_{X\mu}(x) {\cal J}_{X\nu}  \ket{0}\nn
\ee
where ${\cal J}_X$ denotes any choice of normalised generator, e.g. ${\cal J}_X = {\cal J}_X^1$; the exact choice doesn't matter precisely because it's a symmetric space. We use a similar convention for ${\cal J}_H$. The potential \eqn{potj} can then be written as 
\be
V(U) = -\frac{g_2^2}{2}\mes\Delta^{\mu\nu}(x) \sum_\alpha \Big[ &&\!\!  \Tr (U^\dagger G^\alpha U)_H^2\,  \bra{0}  {\mathcal{J}}_{H\mu}(x) {\mathcal{J}}_{H\nu} (0)\ket{0}  \nn\\ &&+\  \Tr (U^\dagger G^\alpha U)_X^2\,   \bra{0}  {\mathcal{J}}_{X\mu}(x) {\mathcal{J}}_{X\nu} (0)\ket{0}   \Big]\label{potj2}
\ee
We can further simplify this using
\be \Tr (G^\alpha)^2 =  \Tr( U^\dagger G^\alpha U )^2 &=& \Tr ( (U^\dagger G^\alpha U)_X + (U^\dagger G^\alpha U)_H)^2 \nn \\ &=& 
\Tr  (U^\dagger G^\alpha U)_X^2 + \Tr (U^\dagger G^\alpha U)_H^2 \nn\ee
Both terms in the potential \eqn{potj2} can then be written in terms of $\Tr  (U^\dagger G^\alpha U)_X^2$, giving
\be
V(U) = V_0 + \frac{g_2^2}{2}\frac{f_\pi^2 M^2}{4\pi}\sum_{\alpha}\Tr (U^\dagger G^\alpha U)_X^2\label{goodpot}\ee
where $V_0$ is independent of $U$. Here we've introduced $f_\pi$, the characteristic energy scale associated to chiral symmetry breaking, defined in the usual manner as
\be
\bra{0}{\cal J}^{i}_{X\mu}\ket{\pi^j} = i  f_\pi p_\mu \delta^{ij}\nn
\ee
The mass scale  $M^2$ in \eqn{goodpot}  is given in terms of broken and unbroken current by
\be
M^2 = \frac{4\pi}{f_\pi^2}\mes \Delta^{\mu\nu}(x)\bra{0}{\cal J}_{H\mu}(x){\cal J}_{H\nu} (0)- {\cal J}_{X\mu}(x){\cal J}_{X\nu} (0)\ket{0}
\label{m2}\ee
Importantly, $M^2$ can be shown to be positive \cite{peskin,preskill}. We postpone the derivation of this result to Appendix \ref{2ndsec} below.

The expression for the potential \eqn{goodpot} has a particularly elegant interpretation: the group theoretic factor is simply the sum of the $G_2$ gauge boson masses,
\be
\sum_\alpha  \Tr (U^\dagger G^\alpha U)_X^2 \sim \sum\left(\text{gauge boson mass}\right)^2 \nn
\ee
We see that, with $M^2>0$, the minimum of the potential $V(U)$ occurs when the gauge group is broken the least, in the sense that the sum of the gauge boson masses is smallest. 

In practice, life is simplest if we are able to pick the reference state $\ket{0}$ to be a local minimum. For this to be the case, the  generators $G^\alpha$ of $G_2\subset K$ must obey a number of properties. To see this, we parameterise the vacua in the neighbourhood of $\ket{0}$ by
\be U(\rho) = \exp(i \rho_i X^i)\nn\ee
To leading order in $\rho$, the potential \eqn{goodpot} then reads
\be V(\boldsymbol{\rho}) = V_0 + \frac{g_2^2}{2}\frac{f_\pi^2 M^2}{4\pi}\sum_{\alpha}\Big(\Tr \left(G^\alpha_X)^2 + i \rho_i \Tr (G^\alpha_X\left[G^\alpha, X^i\right]_X\right) + \ldots\Big)\nn\ee
For $\rho=0$ to be an extremum of $V$, we need 
\be
\frac{\partial}{\partial \rho_i}V(0) &\sim& \sum_\alpha\Tr G^\alpha_X\left[G^\alpha,X^i\right]_X\nn\\
&=&  \sum_\alpha  \Tr G^\alpha_X\left[G^\alpha_H,X^i\right]\nn\\
&=&  \sum_\alpha \Tr X^i \left[G^\alpha_H, G^\alpha_X\right] =0\nn
\ee
where the second equality follows from the fact that, for  $K/H$  a symmetric space, $\left[H,H\right]\sim i H$ and $\left[H,X\right]\sim i X$, and $\left[X,X\right] \sim i T$. The third equality is of course the cyclic property of trace.  We learn that the reference vacuum $\ket{0}$ is a stationary point of $V$ provided that
\be
\left[G^\alpha_H,G^\alpha_X\right] = 0 \ \ \ \mbox{for each}\ \alpha\ \ \ (\mbox{no sum})\label{keepstill}\ee
Next we must ensure that $\ket{0}$ is a local minimum, as opposed to a maximum or saddle point. For this, we must compute the Hessian of $V$. In a mass-diagonal basis for the broken generators $X^i$, one can show that the mass eigenstates are given by
\be m_X^2 = \frac{g_2^2M^2}{4\pi}\sum_{\alpha} \Big[\Tr\left[G^\alpha_H,\left[G^\alpha_H,X\right]\right]X - \Tr\left[G^\alpha_X,\left[G^\alpha_X,X\right]\right]X\Big]\label{pgb}\ee
This combination will show up regularly in what follows; we denote it as
\be m_X^2 = \frac{g_2^2M^2}{4\pi}\sum_{\alpha}\mathcal{C}_X\left(G^\alpha\right) \label{pgb2}
\ee
We see that we have a local minimum only if $m^2_X>0$ for each of the pseudo-Goldstone bosons $X$.  In contrast, if there is any direction with $m^2_X<0$ then there is a tachyonic mode which destabilises the would-be vacuum. 

In fact, life is not quite as simple as we have described. We will encounter a number of situations in which the leading order result \eqn{pgb} gives $m^2_X=0$ for some pseudo-Goldstone boson $X$, even though there is no symmetry protecting the mass. In this case, we must  work harder and look to the second-order terms.

\subsection{Second Order Corrections to the Potential}\label{2ndsec}

To compute the second order corrections to the masses of pseudo-Goldstone bosons, we need a little bit of non-perturbative information. Fortunately, this information is available in the form of sum rules, first derived by Weinberg \cite{sumrule}. Moreover, this machinery is precisely what's required to prove that $M^2$, defined in \eqn{m2}, is positive definite. We now review this, following \cite{peskin,preskill}. 

The spectral function $\rho_H(s)$, corresponding to the unbroken currents ${\cal J}_H$ is defined by
\be
\bra{0}{\cal J}_H^\mu(x){\cal J}_H^\nu(0)\ket{0} = \int_0^\infty \dd s\int\frac{\dd^4 p}{(2\pi)^4} \, \rho_H(s)\frac{-i e^{-i p\cdot x}}{p^2-s+\ii \epsilon}\left(\eta^{\mu\nu}-\frac{p^\mu p^\nu}{p^2}\right)\nn\ee
For the broken currents ${\cal J}_X$, the corresponding spectral function $\rho_X$ has an extra term,
\be
\bra{0} {\cal J}_X^\mu(x){\cal J}_X^\nu(0)\ket{0} = \int_0^\infty \dd s\int\frac{\dd^4 p}{(2\pi)^4} \,\frac{-ie^{-i p\cdot x}}{p^2-s+\ii \epsilon} \left[\rho_X(s)\left(\eta^{\mu\nu}-\frac{p^\mu p^\nu}{p^2}\right)-f_\pi^2 \delta(p^2) p^\mu p^\nu\right]\nn
\ee
Importantly, the spectral functions obey a number of sum rules \cite{sumrule}, 
\be
\int_0^\infty \dd s \left(\rho_H(s)-\rho_X(s)\right) &=& 0 \nn\\
\int_0^\infty \frac{\dd s}{s}\left(\rho_H(s)-\rho_X(s)\right) &=& f_\pi^2\label{summy}
\ee
The mass $M^2$ can be written in terms of the two spectral functions  (see, for example, \cite{preskill}) as
\be
M^2 = \frac{3}{4\pi f_\pi^2} \int_0^\infty \dd s \ \log \frac{s_0}{s}\left(\rho_H(s)-\rho_X(s)\right)\nn
\ee
where $s_0$ is a regularisation scale. To simplify this further, we must assume that the spectral functions are dominated by the lowest lying mesons, and are correspondingly approximated by delta-functions
\be
\rho_H \simeq \lambda_1^2 \,\delta(s-M_H^2)\qquad {\rm and} \qquad \rho_X \simeq \lambda_2^2\,\delta (s-M_X^2)\nn
\ee
Here  $M_H$ and $M_X$ are the masses of the lowest-lying spin-1 mesons coupled to the unbroken and broken currents respectively\footnote{For orientation, in QCD with $N_f=2$ flavours, the broken and unbroken generators arise from the chiral symmetry breaking pattern $SU(2)_L\times SU(2)_R\rightarrow SU(2)_{\rm diag}$. Here the $\rho$ meson, with $M_H=770\ {\rm MeV}$ couples to ${\cal J}_H$ while the $a_1$ meson, with mass $M_X=1260\ {\rm MeV}$ couples to ${\cal J}_X$.}
 and $\lambda_1,\lambda_2$ are the strengths of the couplings. The two sum rules \eqn{summy} then tell us that
 $\lambda_1^2 \simeq \lambda_2^2 = \lambda$ from the second equation and
\be \frac{1}{M_H^2} - \frac{1}{M_X^2} = \frac{f_\pi^2}{\lambda^2}\nn\ee
%
%
Since the right-hand side is positive, we learn that $M_X^2 > M_H^2$. This is sufficient to guarantee positivity of $M^2$ which can be related to the meson masses as
\be M^2 = \frac{3\lambda^2}{4\pi f_\pi^2} \log\left(\frac{M_X^2}{M_H^2}\right)\nn\ee

The machinery of spectral functions is also needed to get an expression for the second-order correction to the masses of the pseudo-Goldstone bosons. The essence of the idea is simple: the masses $M^2$ are computed using the gapless gauge boson propagator $\Delta^{\mu\nu}(x)$ in \eqn{m2}. However,  the condensate partially breaks the gauge group $G_2$, giving some of the gauge bosons a mass.  For these gauge bosons, the propagator should be replaced by the massive propagator.

Here for simplicity, we assume that each of the massive $G_2$ gauge bosons has the same mass, which we denote as $\mu^2$. 
(In the examples of Appendix \ref{bsec}, this is too naive and there will be gauge bosons with different masses. This adds an extra complication, but here we deal with just the simplest case.) The mass $\mu^2$  changes the propagator of the gauge boson and, correspondingly, shifts the mass $M^2$ to $M_\mu^2$, which we will compute shortly. Note that the mass $\mu^2$ will be proportional to $g_2^2$, meaning that $M_\mu^2$ differs from $M^2$ only at order  $g_2^4$.

With this correction, the mass of the pseudo-Goldstone bosons \eqn{pgb2} becomes
\be
m^2_X = \frac{g_2^2}{4\pi}\left(M^2\!\!\!\!\sum_{\text{unbroken}}\!\!\!\!\mathcal{C}_X(G^\alpha)  + M^2_{\mu}\!\sum_{\text{broken}}\!\!\mathcal{C}_X(G^\alpha)\right)\nn\ee
Our interest lies in those pseudo-Goldstone bosons whose mass $m_X^2$ vanishes at leading order. For these, it must be the case that $\sum_{\rm unbroken} {\cal C}_X(G^\alpha) = \sum_{\rm broken} {\cal C}_X(G^\alpha)$, so we can write
\be
m_X^2= \frac{g_2^2}{4\pi}\sum_{\text{unbroken}}\mathcal{C}_X(G^\alpha )\left(M^2-M^2_\mu\right)\label{notzero}
\ee
Note, however, that for an unbroken generator $G^\alpha$ we have, by definition, $G^\alpha_X=0$ (since $G_X^\alpha$ is the projection onto the broken part). Using the definition of ${\cal C}_X(G^\alpha)$ in \eqn{pgb} and \eqn{pgb2}, we see that 
\be \sum_{\rm unbroken}\mathcal{C}_X(G^\alpha ) >0\nn\ee
This is the key to showing that the second order correction to the mass terms is positive. (It is also the step that needs revisiting when the broken gauge bosons have different masses.) Invoking the spectral representation, the mass \eqn{notzero} can be written as
\be
  m^2_X &=& \frac{3g_2^2}{(4\pi)^2 f_\pi^2}\left(\sum_{\text{unbroken}}\mathcal{C}_X(G^\alpha)\right)\int d s\ \left(\log\frac{s_0}{s} - \frac{s}{s-\mu^2}\log\frac{\mu^2}{s}\right)\left(\rho_H-\rho_X\right)\nn\\
      &\approx& \frac{3g_2^2}{(4\pi)^2 f_\pi^2}\left(\sum_{\text{unbroken}}\mathcal{C}_X(G^\alpha)\right)\int d s\ \frac{\mu^2}{s}\log\frac{s}{\mu^2} \left(\rho_H(s)-\rho_X(s)\right)\nn\\
  &\approx& \frac{3g_2^2\lambda^2}{(4\pi)^2 f_\pi^2}\left(\sum_{\text{unbroken}}\mathcal{C}_X(G^\alpha)\right) \left(\frac{\mu^2}{M_H^2} \log\frac{M_H^2}{\mu^2}- \frac{\mu^2}{M_X^2} \log\frac{M_X^2}{\mu^2} \right)\label{masscorrection}
\ee
The fact that this is positive definite follows once again from the observation that $M_X^2> M_H^2$. 
This ensures that those pseudo-Goldstone bosons that remain massless at leading order receive a positive mass at the next order. Note also that the gauge boson mass is of order $\mu^2 \sim g_2^2 f_\pi^2$, ensuring that this mass $m_X^2$ is indeed of order $g_2^4$ as expected.

As we stressed above, this calculation assumed that the massive $G_2$ gauge bosons have a common mass $\mu^2$. This allowed us to write the second-order correction to the massless pseudo-Goldstone bosons as \eqn{notzero}. Below we will meet situations in which this step needs revisiting, and the positivity of the mass correction is no longer so straightforward. Nonetheless, we will see that the positivity remains.

\section{Examples}\label{bsec}

We now apply the results of Appendix \ref{asec} to the models considered in the bulk of the paper. 

\subsection{Vacuum Alignment for $SU(2)\times SU(3)$}\label{SMalign}

We start by applying the ideas above to the chiral gauge theory with gauge group $G=SU(2) \times SU(3)$, coupled to $N_f$ generations of fermions. For now, we include neither hypercharge nor Yukawa interactions.  This is the theory described in Section \ref{nfmore}. 

When $\Lambda_{\rm strong}\gg \Lambda_{weak}$, so the strong force dominates, the original chiral symmetry breaking gives rise to a moduli space ${\cal M}_0 = [SU(2N_f)_L\times SU(2N_f)_R]/SU(2N_f)$. In this case, there is no calculation to do: each point in ${\cal M}_0$ breaks the $SU(2)$ gauge group completely. As described in the main text, the  true moduli space of the theory is ${\cal M}_{\rm strong}$ defined in \eqn{itsstrong}. We have $\dim {\cal M}_0 - \dim {\cal M}_{\rm strong}= 3$, with this difference accounted for by the Higgs mechanism which means that three pions are eaten when $SU(2)$ is broken. This simple counting means that there are no pseudo-Goldstone bosons in this case and no  potential over ${\cal M}_0$ is generated. 

The regime $\lweak\gg\lstrong$ is more involved. When the $SU(2)$ gauge group becomes strong, the resulting condensate \eqn{weakly} allows for a number of different symmetry breaking patterns. These include $SU(3)_{\rm strong} \rightarrow SU(2)_{\rm strong}$, and $SU(3)_{\rm strong}\rightarrow \varnothing$.
We  show here that the former symmetry breaking pattern is a (local) minimum of the potential.  In Appendix \ref{unstablesec} we show that putative vacua in which $SU(3)_{\rm strong}$ is completely broken have a tachyon and are unstable. 

We denote the fermions as
\be
\psi_{mi} = \left(q^1_{Li},q^3_{Li},q^2_{Li},l_{Li}\right)\ \ \ \ {\rm with}\ m=1,2,3,4\ {\rm and}\ i=1,\ldots N_f\nn\ee
(Note that the colour components $q^3$ and $q^2$ are exchanged compared to the main text. This doesn't change the conclusions, but makes some of the generators below a little simpler.) If we ignore the $SU(3)_{\rm strong}$ gauge fields, we have a moduli space of vacua given by
\be {\cal M}_0 = K/H =\frac{SU(4N_f)}{Sp(2N_f)}\nn\ee
We now turn on the $SU(3)_{\rm strong}$ gauge fields. We will show that the flavour diagonal ground state
\be \langle \psi_{mi} \cdot \psi_{nj}\rangle\sim   \tiny{\left(\begin{array}{cccc}
&-1  & & \\
1 & & & \\
 &   & & -1 \\
  & &1  & 
\end{array}\right)}_{mn} \delta_{ij}\label{fvac}\ee
is a minimum of the resulting potential.

It is trivial to show that this vacuum is an extremum of the potential, with the generators obeying \eqn{keepstill}; this follows from the flavour-diagonal nature and the fact that there is no vacuum alignment problem for a  single generation. It remains to show that the masses \eqn{pgb} of the pseudo-Goldstone bosons are non-tachyonic. For this, we will need explicit expressions for the generators of $G_2 = SU(3)_{\rm strong}$ and $X\in SU(4N_f)/Sp(2N_f)$.

First the gauge generators. Since the vacuum \eqn{fvac} breaks $SU(3)_{\rm strong}\longrightarrow SU(2)_{\rm strong}$, it makes sense to classify the generators in terms of their representation under $SU(2)_{\rm weak}$. They decompose as 
${\bf 8} \longrightarrow {\bf 3}\oplus 2({\bf 2}) \oplus {\bf 1}$. The triplet 
\be
G^{\alpha}_{{\bf 3}} = \frac{1}{\sqrt{2N_f}} \sigma^{\alpha} \otimes\begin{pmatrix}1&0\\0&0\end{pmatrix}\otimes\textbf{1}_{N_f} \ \ \ \  \alpha = 1,2,3\nn\ee
are the generators of the unbroken $SU(2)_{\rm strong}$ where, here and below, the generators are normalised as ${\rm Tr}\,G^\alpha G^\beta = \delta^{\alpha\beta}$. The two pairs of generators transforming in the doublet are
\be
  G^{\alpha}_{{\bf 2}} &\in&  \Bigg\{\frac{1}{\sqrt{2N_f}} {\tiny\left(
                         \begin{array}{cc|cc}
                           &1  & & \\
                           1& & & \\
                           \hline
                           &   &\phantom{0} & \\
                           & & &\phantom{0}
\end{array}\right)} \otimes\textbf{1}_{N_f}, \frac{1}{\sqrt{2N_f}}{\tiny \left(\begin{array}{cc|cc}
                                                                        &-i  & & \\
                                                                        i& & & \\
                                                                        \hline
                                                                        &   &\phantom{0} & \\
                                                                        & & &\phantom{0}
\end{array}\right)}\otimes\textbf{1}_{N_f} \Bigg\}\ , \nn\\
&& \Bigg\{\frac{1}{\sqrt{2N_f}}{\tiny\left(\begin{array}{cc|cc}
&  & & 0 \\
& &1 & \\
\hline
 & 1  & & \\
 0 & & &
\end{array}\right)}\otimes\textbf{1}_{N_f}, \frac{1}{\sqrt{2N_f}} {\tiny \left(\begin{array}{cc|cc}
&  & & 0\\
& &-i & \\
\hline
 & i & & \\
 0  & & & 
\end{array}\right)}\otimes\textbf{1}_{N_f}\Bigg\}\nn\ee
Finally the singlet is given by
\be G_{{\bf 1}} = \frac{1}{\sqrt{6N_f}}\diag \left(1,-2,1,0\right)\otimes\textbf{1}_{N_f}\nn
\ee
All gauge generators are singlets under the unbroken $SO(N_f)$ flavour group.

Next, the (pseudo)-Goldstone modes. Under the original symmetry breaking $SU(4N_f)\longrightarrow Sp(2N_f)$, the broken generators
transform in the traceless antisymmetric rank-2 tensor representation of
$Sp(2N_f)$, denoted by $\mathcal{A}$. We have
\be \text{dim}\,\mathcal{A} = (2N_f-1)(4N_f+1)\nn\ee
After gauging $SU(3)_{\rm colour}$, the global group $H=Sp(2N_f)$ is broken to
\be Sp(2N_f) \longrightarrow SU(2)_{{\rm strong}}\times SO(N_f)\times U(1)_{\hat{V}}\nn\ee
Under this decomposition, the  branching rule for the anti-symmetric representation ${\cal A}$ reads
\be \mathcal{A} \longrightarrow 2\left({\bf 2},{\bf 1}\right)_0\oplus \left({\bf 1}, {\bf 1}\right)_0\oplus 2\left({\bf 2}, S\right)_0\oplus \left({\bf 1}, S\right)_0\oplus
\left({\bf 1}, S\oplus A\right)_0\oplus \left({\bf 3},A\right)_0\oplus \left({\bf 2}, A\right)_{\pm 3}\oplus \left({\bf 1}, A\right)_{\pm 6}\nn\ee
where $S$ and $A$ are the traceless symmetric and the antisymmetric rank-2 tensor representation of $SO(N_f)$, respectively; they have dimensions
\be
  {\rm dim}\,S = \frac{N_f}{2}(N_f+1) - 1\ \ \ {\rm and}\ \ \ 
  {\rm dim}\,A = \frac{N_f}{2}(N_f-1)\nn
\ee
The five generators sitting in singlet representations of $SO(N_f)$, namely $2({\bf 2},{\bf 1})_0\oplus ({\bf 1}, {\bf 1})_0$ are the only generators that remain in the case $N_f=1$. These are the five Goldstone modes that become the longitudinal modes of the massive gauge bosons as $SU(3)_{\rm strong}\rightarrow SU(2)_{\rm strong}$. 

We need explicit forms for the remaining generators. This is aided by the observation that, for the condensate \eqn{fvac} with symmetry breaking pattern $SU(4N_f)\rightarrow Sp(2N_f)$, the unbroken and broken generators take the form
\be H = \left(\begin{array}{c|c} A & B \\ \hline B^T & -A^T\end{array}\right)\ \ \ {\rm and}\ \ \ X = \left(\begin{array}{c|c} C & D \\ \hline D^T & C^T\end{array}\right)\nn\ee
with $A$ Hermitian, $B$ symmetric, $C$ traceless Hermitian and $D$ anti-symmetric. In their full glory, the broken generators are:
\begin{itemize}
\item The pair of $({\bf 2},S)_0$ representations are generated by matrices of the form
\be
X_{({\bf 2},S)} &= \frac{1}{2\sqrt{N_f}} {\tiny \left(\begin{array}{cc|cc}
           &z& & \\
z^* &   & & \\
                  \hline
& & & z^*\\
 & &z &  
 \end{array}\right)} \otimes S \ \ \ {\rm and}\ \ \ 
    \frac{1}{2\sqrt{N_f}}{\tiny\left(\begin{array}{cc|cc}
&  & &-z^* \\
& & z^*& \\
                  \hline
 & z  & & \\
 -z & & &
\end{array}\right)} \otimes S \nn
\ee
with $S$ a traceless, symmetric matrix and $z\in \{1,i\}$.
\item The representation $({\bf 1},S)_0$ is generated by matrices of the form
\be
    X_{({\bf 1},S)} = \frac{1}{2\sqrt{N_f}}\text{diag}\left(1,-1,1,-1\right)\otimes S\nn
\ee
\item The representation $({\bf 1},S\oplus A)_0$ is generated by matrices of the form
\be
    X_{({\bf 1},S \oplus A)} = \frac{1}{\sqrt{2N_f}}\text{diag}\left(0,1,0,1\right)\otimes L\nn
\ee
with $L$ a traceless, Hermitian matrix. 
\item The representation $({\bf 3},A)_0$ is generated by matrices of the form
\be
    X_{({\bf 3},A)} = \frac{1}{\sqrt{2N_f}}\sigma^i\otimes\begin{pmatrix} 1& 0\\ 0&0\end{pmatrix}\otimes A 
\nn\ee
with $A$ an anti-symmetric Hermitian matrix.
\item The pair of representations $({\bf 2},A)_{\pm 3}$ is generated by matrices of the form
\be
                               X_{({\bf 2},A)} &=  \frac{1}{2\sqrt{N_f}} {\tiny\left(\begin{array}{cc|cc}
           &z& & \\
z^* &   & & \\
                  \hline
& & & -z^*\\
 & & -z &  
                           \end{array}\right)}\otimes A \ \ \ {\rm and}\ \ \ 
                           \frac{1}{2\sqrt{N_f}} {\tiny \left(\begin{array}{cc|cc}
&  & &z^* \\
& & z^*& \\
                  \hline
 & z  & & \\
 z & & &
                              \end{array}\right)}\otimes A\nn
                               \ee
again with $z\in \{1,i\}$.
\item Finally, the pair of representations $({\bf 1}, A)_{\pm 6}$ is generated by matrices of the form
\be
    X_{({\bf 1},A)} = \frac{1}{\sqrt{2N_f}}{\tiny \left(\begin{array}{cc|cc}
\phantom{0} &\phantom{0} &\phantom{0} &\phantom{0} \\
& & &z^* \\
                  \hline
\phantom{0} &\phantom{0} &\phantom{0} &\phantom{0} \\
 &z  & &\end{array}\right)}\otimes A\nn
\ee
\end{itemize}
Each of the $N_f\times N_f$ matrices above is normalised such that
\be \Tr L^2 = \Tr S^2 = \Tr A^2 = N_f\nn\ee
ensuring that the generators have normalisation $\Tr X^2 = 1$. 
Note that in the basis given above, $({\bf 2},A)_{\pm 3}$ and $({\bf 1}, A)_{\pm 6}$
are not $U(1)_{\hat{V}}$ diagonal, but they can be made diagonal in the full
$SU(2)_{{\rm strong}}\times SO(N_f)\times U(1)_{\hat{V}}$ under a unitary change of basis.

With these explicit expressions, it is now a simple matter to compute the masses of the various generators using \eqn{pgb}. We find three of the representations have mass,
\be
  m^2_{({\bf 3},A)} = \frac{g_s^2 M^2}{\pi N_f},\quad m^2_{({\bf 2},A)} = \frac{g_s^2M^2}{2\pi N_f}, \quad m^2_{({\bf 1},A)} = \frac{g_s^2M^2}{6\pi N_f}\label{massa1}\ee
with $g_s$ the gauge coupling of $SU(3)_{\rm strong}$. 
Each of these is positive, as is required for a stable ground state.
The remaining three generators  are massless
\be 
  m^2_{({\bf 2},S)} &= m^2_{({\bf 1},S)} =
                      m^2_{({\bf 1},S\oplus A)} = 0\label{masszeroa1}
\ee
Of these massless generators, $({\bf 1},S)_0$ and $({\bf 1},S\oplus A)_0$ are neutral under the $SU(2)_{\rm strong}$ gauge group and so we do not expect them to receive any further corrections. Indeed, these generators correspond to the exact Goldstone bosons of the theory. We can confirm this with some simple counting, 
\be
\left(\frac{1}{2}N_f(N_f+1) -1\right) + (N_f^2-1) = \frac{3}{2}N_f^2 +\frac{N_f}{2} - 2\nn
\ee
which coincides ${\rm dim}\, \mathcal{M}^\prime_{{\rm weak}}$ defined in \eqn{mprimeweak}, the expected number of exact Goldstone modes. 

This leaves us with the fate of the pair of  $({\bf 2},S)_0$ representations unaccounted for. These are not exact Goldstone bosons, so we expect that the vanishing of the mass is an artefact of working to leading order in perturbation theory; we must look to second order to see if the resulting mass-squared is positive or negative.

\subsubsection*{Second Order Corrections}

We now adapt the results of Appendix \ref{2ndsec} to determine the second order correction to the $({\bf 2},S)_0$ states. As explained previously, the relevant physics comes from taking into account the mass splitting of the broken $SU(3)_{\rm strong}$ gauge generators. One key difference with the results of Appendix \ref{2ndsec} is that now these gauge bosons have different masses. 

It will be useful to describe the general case, in which the  broken gauge generators sit in more than one irreducible representations of the unbroken gauge group $\tilde{G}$,
\be {\bf r}_{{\rm broken}} = \bigoplus_{i\in I} {\bf r}_i \nn\ee
The masses of gauge bosons in each representation ${\bf r}_i$ will, in general, differ. We  denote these masses as $\mu_i^2$. The analysis of Appendix \ref{2ndsec} then proceeds, with the final result \eqn{masscorrection} replaced by
\be 
 m_X^2 \approx  - \frac{3g_s^2\lambda^2}{(4\pi)^2 f_\pi^2} \sum_{i\in I}  \left(\sum_\alpha \mathcal{C}_X(G_{{\bf r}_i}^\alpha)\right) \left(\frac{\mu_i^2}{M_H^2} \log\frac{M_H^2}{\mu_i^2}- \frac{\mu_i^2}{M_X^2} \log\frac{M_X^2}{\mu_i^2} \right)\label{newcorrection}\ee
Note that, in contrast to \eqn{masscorrection}, we are now summing over the {\it broken} generators, rather than the unbroken generators. This is compensated by the overall minus sign that sits in front. The fact that $M_X^2> M_H^2$ ensures that the log terms are positive definite. To ensure stability, we now need that the group theory factor is negative, to cancel the overall minus sign.

Our example of interest has  $\tilde{G}=SU(2)$ and the broken generators sit in ${\bf 1}\oplus {\bf 2}\oplus {\bf 2}$. The masses of the corresponding gauge bosons are given by
\be \mu_{{\bf 1}}^2 = \frac{2}{3} g_s^2f_\pi^2\ \ \ {\rm and} \ \ \ \mu_{{\bf 2}}^2 = \frac{1}{2} g_s^2 f_\pi^2\nn\ee
We should then apply \eqn{newcorrection} to the worrisome pseudo-Goldstone mode $X=({\bf 2}, A)$. Happily it turns out that there are no tricky cancellations between group theory factors; instead one finds 
\be \sum_{\alpha=1}^4 {\cal C}_X (G^\alpha_{\bf 2}) =-1 \ \ \ {\rm and}\ \ \  {\cal C}_X(G_{\bf 1}) = -\frac{1}{2}\nn\ee
The fact that each of these is negative, means that they both contribute positively to the mass $m_X^2$. We see the vacuum \eqn{fvac} remains stable at second order.

\subsection{Unstable Vacua}\label{unstablesec}

We have shown that the flavour-diagonal condensate  \eqn{fvac} is a local minimum of the potential. We have not, however, shown that it is global minimum.

This, it appears, is a challenging problem. The moduli space is large, and there may be many saddle points. However, the simple observation that the potential   \eqn{goodpot} is proportional to the sum of the W-boson masses means that those ground states which break the gauge symmetry the least are favoured. For this reason, it seems likely that the flavour-diagonal vacuum \eqn{fvac} is, in fact, the true ground state of the system.

In this appendix, we give some calculations to back up this intuition. We have not been able to find other local minima of the potential. Instead, we will show that  a number of obvious candidates for ground states have tachyonic modes and so are unstable.  We work with $N_f=2$ and give two examples of putative ground states, each with different symmetry breaking patterns, which turn out to be saddle points. 

For the first example, consider a condensate of the form
\be \langle q^a_{1\,L} \cdot q^b_{2\,L}\rangle\sim\Lambda^3_{{\rm weak}} \delta^{ab}\ \ \ ,\ \ \ \langle l_{1\,L}\cdot l_{2\,L}\rangle \sim \Lambda^3_{{\rm weak}} \label{unstablevac}\ee
where $a=1,2,3$ is the $SU(3)_{\rm strong}$ colour index, and the 1,2 labels on the quarks and leptons refer to flavour. With such a condensate, the gauge groups breaks as
\be SU(3)_{{\rm strong}} \longrightarrow SO(3)_{{\rm strong}} \nn\ee
If we choose to order the 8 Weyl spinors as $\psi = \left(q^1_{1\,L}, q^2_{1\,L}, q^3_{1\,L},l_{1\,L}|q^1_{2\,L}, q^2_{2\,L}, q^3_{2\,L},l_{2\,L}\right)^{{\rm T}}$, then the  $SU(3)_{{\rm strong}}$ generators act as
\be
G^{\alpha} = \frac{1}{2}\left(\begin{array}{cc|cc}
\lambda^{\alpha}&  & &  \\
& 0 & & \\
\hline
 &  & \lambda^{\alpha}& \\
  & & & 0
 \end{array}\right)\nn
\ee
with  $\lambda^\alpha$ are the usual $3\times 3$ Gell-Mann matrices. It is straightforward to show that the broken and unbroken generators obey \eqn{keepstill}, so this condensate is at least a saddle point of the potential. However, one finds that this condensate has a higher energy than \eqn{fvac}. 
%
More importantly, there are also tachyonic modes. One of this is associated to the would-be Goldstone bosons transforming in the octet of $SU(3)_{{\rm strong}}$
\be
X_{{\bf 8}}^\alpha = \frac{1}{2}\left(
  \begin{array}{cc|cc}
    \lambda^{\alpha}&  & &  \\
                    & 0 & & \\
    \hline
                    &  & \lambda^{\alpha\, {\rm T}}& \\
                    & & & 0
  \end{array}\right)\nn
\ee
The mass matrix for these 8 generators can be found using \eqn{pgb}; it is diagonal, and given by
\be m^2_{\alpha \beta} = -\frac{ g_s^2 M^2}{2\pi} \diag\left(0,1,0,0,1,0,1,0\right) \nn\ee
The overall minus sign means that this vacuum is unstable.

As a second example, we consider the condensate that arises in the single flavour case, but now with two  flavours which align differently within the $SU(3)_{\rm strong}$ gauge group. For the first generation we take,
\be
  \langle q^1_{1\,L}\cdot q^2_{1\,L}\rangle\sim \Lambda^3_{{\rm weak}} \ \ \ ,\ \ \ 
  \langle l_{1\,L}\cdot q^3_{1\,L}\rangle\sim \Lambda^3_{{\rm weak}}
  \label{andone}\ee
which picks out the $a=3$ colour direction. Meanwhile, for the second generation we take
\be
  \langle q^1_{2\,L}\cdot q^3_{2\,L}\rangle\sim \Lambda^3_{{\rm weak}} \ \ \ ,\ \ \ 
  \langle l_{2\,L}\cdot q^2_{2\,L}\rangle\sim \Lambda^3_{{\rm weak}}
  \label{andtwo}\ee
which picks out the $a=2$ colour direction. The combination of both condensates breaks the $SU(3)_{\rm strong}$ gauge group completely. 

To compute the masses, it is simplest to note that the condensate is related to \eqn{fvac} by a permutation of the $\psi$ components. There are two ways to proceed; we could fix the action of the gauge generators $G^\alpha$ on $\psi$, in which case the permutation acts as conjugation on the broken generators $X$ defined in Appendix \ref{SMalign}. Alternatively, we could fix the action of the unbroken generators, in which case the permutation acts by conjugation of $G$. In either case, a simple calculation shows that the condensates \eqn{andone} and \eqn{andtwo} are indeed a saddle point of the potential, but with energy higher than both the local minimum \eqn{fvac} and the unstable vacuum \eqn{unstablevac}.

It is a little more involved to demonstrate that the condensates \eqn{andone} and \eqn{andtwo} are unstable and we refrain from giving all the details. . Because the gauge group is broken completely,  
many of the broken generators $X$ given in Appendix \ref{SMalign} now mix. Diagonalising the resulting mass matrix, one finds that there are massive modes, massless modes and, crucially, two tachyonic modes. This vacuum is  unstable. 

%
  %
  %
  %

\subsection{Adding Hypercharge}\label{hyperalign}

We now repeat the calculation of Appendix \ref{SMalign} in the presence of a $U(1)_Y$ hypercharge interaction. This corresponds to an additional gauge generator which, in the notation of Appendix \ref{SMalign}, takes the form
\be
G_Y = \frac{1}{2\sqrt{3N_f}}\diag\left(1,1,1,-3\right)\otimes {\bf 1}_{N_f}\nn
\ee
The unfamiliar normalisation factor ensures that this generator obeys  $\Tr G_Y^2 = 1$. 

There is a simple generalisation of the mass formula \eqn{pgb} in which the different gauge generators are summed over, weighted with their gauge couplings. We denote the gauge coupling associated to $U(1)_Y$ as $g_Y$. Then the masses of the pseudo-Goldstone bosons \eqn{massa1} are replaced by
\be
  m^2_{({\bf 3},A)_0} = \frac{g_s^2}{\pi N_f}M^2, \quad m^2_{({\bf 2},A)_{\pm 3}} = \frac{g_s^2 M^2}{2\pi N_f}, \quad m^2_{({\bf 1},A)_{\pm 6}} = (g_s^2+2g_Y^{2})\frac{M^2}{6\pi N_f}\nn\ee
 These modes are not destabilised by the hypercharge interaction. Meanwhile, the massless modes \eqn{masszeroa1} remain massless at leading order,
  \be
  m^2_{({\bf 2},S)_0} &= m^2_{({\bf 1},S)_0} = m^2_{({\bf 1},S\oplus A)_0} = 0\nn
\ee
The $({\bf 1},S)_0$ and $({\bf 1},S\oplus A)_0$ states remain as exact Goldstone bosons as previously. There is, however, a correction to the second-order mass of the $({\bf 2},S)_0$ states due to hypercharge. To see this, we need the usual mixing of the generators  $G_{{\bf 1}}$ and $G_{Y}$ which yield the Z-boson and the photon. The mass-matrix for the corresponding gauge fields is
\be
\mu^2 = \frac{f_\pi^2}{3}\begin{pmatrix} 2g_s^2 & \sqrt{2} g_s g_Y\\ \sqrt{2} g_s g_Y & g_Y^{2}\end{pmatrix}\nn
\ee
which has eigenvalues
\be \mu_Z^2 =  \left(2g_s^2 + g_Y^{2}\right)\frac{f_\pi^2}{3}\ \ \ {\rm and}\ \ \  \mu_{\gamma}^2 = 0 \nn\ee
%
%
%
Correspondingly, the generators mix and take the form
\be G_Z = \frac{1}{\sqrt{2g_s^4 + g_Y^{4}}}\left(\sqrt{2}g_s^2\,G_{{\bf 1}} + g_Y^2G_Y\right),\quad G_\gamma = 
\frac{1}{\sqrt{3}}
\left(\sqrt{2} G_Y- G_{{\bf 1}}\right)\nn
\ee
%
%
%
 Armed with these results, we can now revisit the calculation of Appendix \ref{SMalign}. At leading order, only the mass of $({\bf 1}, A)$ is modified by the hypercharge to
\be
m^2_{({\bf 1},A)} = \frac{(g_s^2+2g_Y^2)M^2}{6\pi N_f}\nn
\ee
While the mass of $({\bf 2}, S)$ is modified by replacing the contribution from $G_{\bf 1}$ in \eqn{newcorrection} 
%
%
by the contribution from the Z-boson, which is
\be
 {\cal C}(G_Z) = -\frac{g_s^4 + 2g_s^2 g_Y^2}{2g_s^4+g_Y^4}\nn
\ee
The group theoretic factor remains negative and the mass remains positive.

The upshot of these short calculations is that  hypercharge does not destabilise the vacuum. Indeed, it is clear from the calculations above why this is: both the chosen vacuum, and the hypercharge, are flavour-diagonal. The analysis of \cite{peskin, preskill}  shows that the flavour-diagonal vacuum is likely to be destabilised only by the introduction of  $U(1)$ gauge symmetry under which different generations carry different charges.

\subsection{Vacuum Alignment for $Sp(r)\times SU(N)$}\label{hardalign}

The analysis of vacuum alignment for $Sp(r)\times SU(N)$ follows that of Section \ref{SMalign}; only the group theory is a little more involved.

Before we turn on the $SU(N)_{\rm strong}$ gauge symmetry, the chiral condensate induces the $K\rightarrow H$ symmetry breaking pattern expected of a pseudo-real representation,
\be
SU((N+1)N_f)\longrightarrow Sp((N+1)N_f/2)\nn
\ee
We now gauge $SU(N)_{\rm strong}$. We  postulate that vacuum is again formed by a flavour-diagonal condensate, under which the  gauge group is broken to
\be SU(N)_{{\rm strong}} \longrightarrow  Sp(\nu)_{{\rm strong}}\ \ \ \ {\rm with}\ \nu = \frac{N-1}{2}\nn
\ee
More generally, as explained in Section \ref{itsnewsec}, the surviving  global symmetry $H$ is broken to 
\be
H=Sp((N+1)N_f/2) \longrightarrow Sp(\nu)_{{\rm strong}} \times SO(N_f) \times U(1)_{\hat{V}}\label{decagain}
\ee
As before, it will prove useful to decompose  all generators into representations of this  unbroken group. 
The calculation is very similar to that in Appendix \ref{SMalign} apart from one complication
 arising from  the presence of the traceless antisymmetric tensor representation
 of $Sp(\nu)_{{\rm strong}}$ which is absent when $\nu = 1$.

 The $SU(N)_{{\rm strong}}$ generators decompose as
\be {\bf ad} \longrightarrow \hat{\mathcal{S}}\oplus\hat{\mathcal{A}}\oplus 2\left(\hat{\mathcal{F}}\right) \oplus {\bf 1}\label{adidas}\ee
where we use the hat to distinguish these  $Sp(\nu)_{\rm strong}$ representations from similar $SO(N_f)$ representations that we will meet below.  
Here $\hat{\mathcal{S}}$ is the symmetric tensor representation with dimensions $\nu(2\nu+1)$, $\hat{\mathcal{A}}$ is the traceless antisymmetric tensor representation with dimensions $(\nu-1)(2\nu+1)$, and $\hat{\mathcal{F}}$ is the fundamental representation $\boldsymbol{2\nu}$. All are singlets under $SO(N_f)$ and neutral under $U(1)_{\hat{V}}$.

The (pseudo)-Goldstone modes again sit in the traceless antisymmetric rank-2 tensor representation ${\cal A}$ of
$Sp((N+1)N_f/2)$. Under \eqn{decagain}, the branching rules are
\be
{\mathcal{A}}\longrightarrow & (\hat{\mathcal{A}},{\bf 1})_0\oplus 2(\hat{\mathcal{F}},{\bf 1})_0\oplus ({\bf 1},{\bf 1})_0\oplus (\hat{\mathcal{A}},S)_0
                               \oplus 2(\hat{\mathcal{F}}, S)_0\oplus ({\bf 1}, S)_0 \oplus ({\bf 1}, S\oplus A)_0\nn\\
  & \;\oplus (\hat{\mathcal{S}}, A)_0\oplus (\hat{\mathcal{F}}, A)_{\pm N}\oplus ({\bf 1}, A)_{\pm 2N}\nn
\ee
Now we can track the fate of each of these representations.

The representations $(\hat{\mathcal{A}},{\bf 1})_0$, $(\hat{\mathcal{F}},{\bf 1})_0$, and $({\bf 1},{\bf 1})_0$, each of which is a singlet under $SO(N_f)$, are eaten by the Higgs mechanism, and absorbed as longitudinal modes of the massive gauge bosons that arise when $SU(N)_{\rm strong}$ is broken to $Sp(\nu)_{\rm strong}$.

The representations  $\left({\bf 1}, S\right)_0$ and $\left({\bf 1},S\oplus A\right)_0$, which are both singlets under the unbroken $Sp(\nu)_{\rm strong}$ gauge group, are exact Goldstone bosons. They parameterise the moduli space ${\cal M}_{\rm weak} = SU(N_f)_q\times SU(N_f)_l/SO(N_f)$ of the theory defined in \eqn{onelastm}. 

The remaining representations are pseudo-Goldstone bosons.  At leading order, we can compute their masses using the formula \eqn{pgb}. The full calculation is a little involved, and here we quote only the final result:
\be
  m^2_{({\bf 1},A)} = \frac{N-1}{2N}\frac{g_s^2M^2}{2\pi N_f} \ \ \ ,\ \ \ 
  m^2_{(\hat{\mathcal{S}},A)} = \frac{g_s^2 M^2}{\pi N_f}\ \ \ ,\ \ \ 
  m^2_{(\hat{\mathcal{F}},A)} = \frac{5N-7}{2(N-1)}\frac{g_s^2 M^2}{4\pi N_f}\nn
\ee
Meanwhile, at leading order, two of the pseudo-Goldstone bosons remain massless,
\be m^2_{({\hat{\mathcal{A}}},S)} &= m^2_{(\hat{\mathcal{F}},S)} = 0\nn\ee
We write these as $m^2_{\cal A}$ and $m_{\cal F}^2$ for short. As in previous examples, to see whether these massless modes will destabilise the vacuum we need to look at higher order. 

The broken gauge generators sit in the $\hat{\cal A}$, $\hat{\cal F}$ and singlet representations in  \eqn{adidas}. The masses of the corresponding gauge bosons are given by
\be \mu^2_{\mathcal{A}} = g_s^2 f_\pi^2 , \quad \mu^2_{\mathcal{F}} = g_s^2\frac{1}{2}f_\pi^2 , \quad \mu^2_{{\bf 1}} = \frac{N+1}{2N} g_s^2f_\pi^2  \nn
\ee
We now use the formula \eqn{newcorrection}; each of the $X=({\hat{\mathcal{A}}},S)$ and $X=(\hat{\mathcal{F}},S)$ pseudo-Goldstone modes get mass at second order, given by
\be 
 m_X^2 \approx  - \frac{3g_s^2\lambda^2}{(4\pi)^2 f_\pi^2} \sum_{i\in \{{\cal A},{\cal F},{\bf 1}\}}  \left(\sum_\alpha \mathcal{C}_X(G_{{\bf r}_i}^\alpha)\right) \left(\frac{\mu_i^2}{M_H^2} \log\frac{M_H^2}{\mu_i^2}- \frac{\mu_i^2}{M_X^2} \log\frac{M_X^2}{\mu_i^2} \right)\nn\ee
Once again, each of the representations contributes a positive contribution to the mass. For the $X=({\hat{\mathcal{A}}},S)$  state, we have
\be \sum_\alpha {\cal C}_{({\hat{\mathcal{A}}},S)} (G^\alpha_{\cal A}) =-(N-5) \ \ \ , \ \ \  \sum_\alpha {\cal C}_{({\hat{\mathcal{A}}},S)} (G^\alpha_{\cal F}) =-\frac{5N-13}{N-3} \ \ \ ,\ \ \ 
 {\cal C}_{({\hat{\mathcal{A}}},S)} (G_{\bf 1})= 0\nn\ee
Note that there is no ${\cal A}$ representation when $N=3$, and these formulae hold only for $N\geq 5$.   Meanwhile, for the $X=(\hat{\mathcal{F}},S)$ state, we have
\be \sum_\alpha {\cal C}_{({\hat{\mathcal{F}}},S)} (G^\alpha_{\cal A}) =-\frac{N-3}{2} \ \ \ , \ \ \  \sum_\alpha {\cal C}_{({\hat{\mathcal{F}}},S)} (G^\alpha_{\cal F}) =-\frac{3N-5}{2(N-1)} \ \ \ ,\ \ \ 
 {\cal C}_{({\hat{\mathcal{F}}},S)} (G_{\bf 1})=-\frac{1}{N-1}\nn\ee
The fact that each of these is negative ensures that the masses of the pseudo-Goldstone bosons are positive and the vacuum is stable. One can further show that the vacuum is not destabilised by the addition of hypercharge.

%

%
%


\newpage

\begin{thebibliography}{99}

\small
\parskip=0pt plus 2pt





\bibitem{wein2}   S.~Weinberg,
  ``{\it Implications of Dynamical Symmetry Breaking: An Addendum},''
 Phys.\ Rev.\ D {\bf 19}, 1277 (1979).
\bibitem{susskind}   L.~Susskind,
  ``{\it Dynamics of Spontaneous Symmetry Breaking in the Weinberg-Salam Theory},''
  Phys.\ Rev.\ D {\bf 20}, 2619 (1979).
  
\bibitem{sam}   S.~Samuel,
  ``{\it The Standard model in its other phase},''
  Nucl.\ Phys.\ B {\bf 597}, 70 (2001)
  [hep-ph/9910559].
  
\bibitem{quigg}  C.~Quigg and R.~Shrock,
  ``{\it Gedanken Worlds without Higgs: QCD-Induced Electroweak Symmetry Breaking},''
  Phys.\ Rev.\ D {\bf 79}, 096002 (2009)
  [arXiv:0901.3958 [hep-ph]].
  
\bibitem{shrock}   Y.~L.~Shi and R.~Shrock,
  ``{\it Dynamical Symmetry Breaking in Chiral Gauge Theories with Direct-Product Gauge Groups},''
  Phys.\ Rev.\ D {\bf 94}, no. 6, 065001 (2016)
  [arXiv:1606.08468 [hep-th]].

\bibitem{af}  L.~F.~Abbott and E.~Farhi,
  ``{\it Are the Weak Interactions Strong?},''
  Phys.\ Lett.\  {\bf 101B}, 69 (1981).

\bibitem{af2}   L.~F.~Abbott and E.~Farhi,
  ``{\it A Confining Model of the Weak Interactions},''
  Nucl.\ Phys.\ B {\bf 189}, 547 (1981).
  

  
  
\bibitem{af3}   M.~Claudson, E.~Farhi and R.~L.~Jaffe,
  ``{\it The Strongly Coupled Standard Model},''
  Phys.\ Rev.\ D {\bf 34}, 873 (1986).




\bibitem{ball}   R.~D.~Ball,
 ``{\it Chiral Gauge Theory},''
  Phys.\ Rept.\  {\bf 182}, 1 (1989).

\bibitem{appel0}  T.~Appelquist, A.~G.~Cohen, M.~Schmaltz and R.~Shrock,
  ``{\it New constraints on chiral gauge theories},''
  Phys.\ Lett.\ B {\bf 459}, 235 (1999)
  [hep-th/9904172].

\bibitem{appel2} T.~Appelquist, Z.~y.~Duan and F.~Sannino,
  ``{\it Phases of chiral gauge theories},''
  Phys.\ Rev.\ D {\bf 61}, 125009 (2000)
  [hep-ph/0001043].
  
  
  
\bibitem{appel3}   T.~Appelquist and R.~Shrock,
  ``{\it Ultraviolet to infrared evolution of chiral gauge theories},''
  Phys.\ Rev.\ D {\bf 88}, 105012 (2013)
  [arXiv:1310.6076 [hep-th]].

\bibitem{ss1} Y.~L.~Shi and R.~Shrock,
  ``{\it Renormalization-Group Evolution of Chiral Gauge Theories},''
  Phys.\ Rev.\ D {\bf 91}, no. 4, 045004 (2015)
  [arXiv:1411.2042 [hep-th]].
 
\bibitem{ss2}  Y.~L.~Shi and R.~Shrock,
  ``{\it Renormalization-Group Evolution and Nonperturbative Behavior of Chiral Gauge Theories with Fermions in Higher-Dimensional Representations},''
  Phys.\ Rev.\ D {\bf 92}, no. 12, 125009 (2015)
  [arXiv:1509.08501 [hep-th]].

\bibitem{ss3}  Y.~L.~Shi and R.~Shrock,
  ``{\it $A_k \bar F$ chiral gauge theories},''
  Phys.\ Rev.\ D {\bf 92}, no. 10, 105032 (2015)
  [arXiv:1510.07663 [hep-th]].
  
\bibitem{stefano}  S.~Bolognesi, K.~Konishi and M.~Shifman,
  ``{\it Patterns of symmetry breaking in chiral QCD},''
  Phys.\ Rev.\ D {\bf 97}, no. 9, 094007 (2018)
  [arXiv:1712.04814 [hep-th]].
  
\bibitem{ryttov}   T.~A.~Ryttov and R.~Shrock,
  ``{\it Ultraviolet to Infrared Evolution and Nonperturbative Behavior of ${\rm SU}(N) \otimes {\rm SU}(N-4) \otimes {\rm U}(1)$ Chiral Gauge Theories},''
  arXiv:1906.04255 [hep-th].



\bibitem{stefano2}   S.~Bolognesi and K.~Konishi,
  ``{\it Dynamics and symmetries in chiral $SU(N)$ gauge theories},''
  arXiv:1906.01485 [hep-th].

\bibitem{anber}   M.~M.~Anber,
  ``{\it Self-conjugate QCD},''
  arXiv:1906.10315 [hep-th].




\bibitem{lattice2}   E.~Poppitz and Y.~Shang,
  ``{\it Chiral Lattice Gauge Theories Via Mirror-Fermion Decoupling: A Mission (im)Possible?},''
  Int.\ J.\ Mod.\ Phys.\ A {\bf 25}, 2761 (2010)
  [arXiv:1003.5896 [hep-lat]].

\bibitem{lattice3}  J. Wang and X. G. Wen, ``{\it Non-Perturbative Regularization of 1+1D Anomaly-Free Chiral
Fermions and Bosons: On the equivalence of anomaly matching conditions and boundary gapping rules,}" arXiv:1307.7480 [hep-lat].
  
\bibitem{lattice4}   J.~Wang and X.~G.~Wen,
  ``{\it A Solution to the 1+1D Gauged Chiral Fermion Problem,}''
  Phys.\ Rev.\ D {\bf 99}, no. 11, 111501 (2019)
  [arXiv:1807.05998 [hep-lat]].
  
 \bibitem{lattice5} J.~Wang and X.~G.~Wen,
  ``{\it A Non-Perturbative Definition of the Standard Models},''
  arXiv:1809.11171 [hep-th].
 
 \bibitem{lattice1}   E.~Eichten and J.~Preskill,
  ``{\it Chiral Gauge Theories on the Lattice,}''
  Nucl.\ Phys.\ B {\bf 268}, 179 (1986).
 
 \bibitem{peskin}   M.~E.~Peskin,
  ``{\it The Alignment of the Vacuum in Theories of Technicolor},''
  Nucl.\ Phys.\ B {\bf 175}, 197 (1980).
  
  
\bibitem{preskill}  J.~Preskill,
  ``{\it Subgroup Alignment in Hypercolor Theories},''
  Nucl.\ Phys.\ B {\bf 177}, 21 (1981).
 
\bibitem{higher1}   D.~Gaiotto, A.~Kapustin, Z.~Komargodski and N.~Seiberg,
  ``{\it Theta, Time Reversal, and Temperature},''
  JHEP {\bf 1705}, 091 (2017)
  [arXiv:1703.00501 [hep-th]].

\bibitem{higher2}   D.~Gaiotto, Z.~Komargodski and N.~Seiberg,
  ``{\it Time-reversal breaking in QCD$_{4}$, walls, and dualities in 2 + 1 dimensions},''
  JHEP {\bf 1801}, 110 (2018)
  [arXiv:1708.06806 [hep-th]].

\bibitem{tanizaki}  Y.~Tanizaki and Y.~Kikuchi,
  ``{\it Vacuum structure of bifundamental gauge theories at finite topological angles},''
  JHEP {\bf 1706}, 102 (2017)
  [arXiv:1705.01949 [hep-th]].

\bibitem{karasik}   A.~Karasik and Z.~Komargodski,
  ``{\it The Bi-Fundamental Gauge Theory in 3+1 Dimensions: The Vacuum Structure and a Cascade},''
  JHEP {\bf 1905}, 144 (2019)
  [arXiv:1904.09551 [hep-th]].
   
\bibitem{scooped}  J.~Berger, A.~J.~Long and J.~Turner,
  ``{\it A phase of confined electroweak force in the early Universe},''
  arXiv:1906.05157 [hep-ph].

\bibitem{thooft}   G.~'t Hooft,
  ``{\it  Naturalness, chiral symmetry, and spontaneous chiral symmetry breaking,}''
  NATO Sci.\ Ser.\ B {\bf 59}, 135 (1980).

 \bibitem{wittensu2}  E.~Witten,
  ``{\it An SU(2) Anomaly,}''
  Phys.\ Lett.\  {\bf 117B}, 324 (1982).
  
\bibitem{tumbling}  S.~Raby, S.~Dimopoulos and L.~Susskind,
  ``{\it Tumbling Gauge Theories},''
  Nucl.\ Phys.\ B {\bf 169}, 373 (1980).
  
\bibitem{light}   S.~Dimopoulos, S.~Raby and L.~Susskind,
  ``{\it Light Composite Fermions},''
  Nucl.\ Phys.\ B {\bf 173}, 208 (1980).
\bibitem{barnone}   I.~Bars and S.~Yankielowicz,
  ``{\it Composite Quarks and Leptons as Solutions of Anomaly Constraints},''
  Phys.\ Lett.\  {\bf 101B}, 159 (1981).

\bibitem{appel}  T.~Appelquist, G.~T.~Fleming and E.~T.~Neil,
  ``{\it Lattice study of the conformal window in QCD-like theories},''
  Phys.\ Rev.\ Lett.\  {\bf 100}, 171607 (2008)
  Erratum: [Phys.\ Rev.\ Lett.\  {\bf 102}, 149902 (2009)]
  [arXiv:0712.0609 [hep-ph]].

\bibitem{su21}
T.~Karavirta, J.~Rantaharju, K.~Rummukainen and K.~Tuominen,
  ``{\it Determining the conformal window: $SU(2)$ gauge theory with $N_f = 4, 6$ and 10 fermion flavours},''
  JHEP {\bf 1205}, 003 (2012)
  [arXiv:1111.4104 [hep-lat]].
\bibitem{su22}  M.~Hayakawa, K.-I.~Ishikawa, S.~Takeda, M.~Tomii and N.~Yamada,
  ``{\it Lattice Study on quantum-mechanical dynamics of two-color QCD with six light flavors},''
  Phys.\ Rev.\ D {\bf 88}, no. 9, 094506 (2013)
  [arXiv:1307.6696 [hep-lat]].
\bibitem{su23} A.~Amato, T.~Rantalaiho, K.~Rummukainen, K.~Tuominen and S.~T\"ahtinen,
  ``{\it Approaching the conformal window: systematic study of the particle spectrum in SU(2) field theory with $N_f=$2,4 and 6.},''
  PoS LATTICE {\bf 2015}, 225 (2016)
  [arXiv:1511.04947 [hep-lat]].

  



  
\bibitem{deconfined}   T.~Senthil, A.~Vishwanath, L.~Balents, S.~Sachdev and M.~P.~A.~Fisher,
  ``{\it Deconfined Quantum Critical Points}", Science 303, 1490 (2004). [arXiv:cond-mat/0311326]



\bibitem{more}   R.~Shrock,
  ``{\it Implications of anomaly constraints in the N(c) extended standard model},''
  Phys.\ Rev.\ D {\bf 53}, 6465 (1996)   [hep-ph/9512430].;   ``{\it Constraints on N(c) in extensions of the standard model},''
  Phys.\ Rev.\ D {\bf 76}, 055010 (2007)
  [arXiv:0704.3464 [hep-th]].


\bibitem{eichten}   E.~Eichten, R.~D.~Peccei, J.~Preskill and D.~Zeppenfeld,
  ``{\it Chiral Gauge Theories in the 1/n Expansion},''
  Nucl.\ Phys.\ B {\bf 268}, 161 (1986).
  
\bibitem{adi}   A.~Armoni and M.~Shifman,
  ``{\it A Chiral SU(N) Gauge Theory Planar Equivalent to Super-Yang-Mills},''
  Phys.\ Rev.\ D {\bf 85}, 105003 (2012)
  [arXiv:1202.1657 [hep-th]].
  
\bibitem{frandsen} M.~Frandsen, T.~Pickup, M.~Teper,
  ``{\it Delineating the Conformal Window}",
  Phys.\ Lett.\ B {\bf 695}, 231 (2011)
  [arXiv:1007.1614 [hep-ph]].
  
\bibitem{sumrule}    S.~Weinberg,
  ``{\it Precise relations between the spectra of vector and axial vector mesons},''
  Phys.\ Rev.\ Lett.\  {\bf 18}, 507 (1967).

\end{thebibliography}
\end{document}